\documentclass[aps,pra,twocolumn,10pt,longbibliography]{revtex4-1}
\usepackage{graphicx} \usepackage{amsmath} \usepackage{bm}
\usepackage{color} \usepackage[T1]{fontenc}

\begin{document}

\title{Impact of spin transfer torque on the write error rate of a
  voltage-torque-based magnetoresistive random access memory}

\author{ Hiroshi Imamura\email{h-imamura@aist.go.jp} and Rie Matsumoto
}

\affiliation{ National Institute of Advanced Industrial Science and
  Technology (AIST), Spintronics Research Center, Tsukuba, Ibaraki
  305-8568, Japan }

\begin{abstract}
  Impact of spin transfer torque (STT) on the write error rate of a
  voltage-torque-based magnetoresistive random access memory is
  theoretically analyzed by using the macrospin model. During the
  voltage pulse the STT assists or suppresses the precessional motion of
  the magnetization depending on the initial magnetization
  direction. The characteristic value of the current density is derived
  by balancing the STT and the external-field torque, which is about
  5$\times$ 10$^{11}$ A/m$^{2}$. The results show that the write error
  rate is insensitive to the STT below the current density of $10^{10}$
  A/m$^{2}$.
\end{abstract}

\maketitle

\section{INTRODUCTION}
Magnetoresistive random access memory (MRAM) is a kind of non-volatile
memory which stores information as stable magnetic states in the
magnetoresistive
devices\cite{Yuasa2013,Apalkov2016,Sbiaa2017,Cai2017}. The stored
information is read by measuring the resistance which strongly depends
on their magnetic states. The MgO-based magnetic tunnel junction (MTJ)
is widely used as a basic element of the MRAM because of its large
read signal\cite{Parkin2004,yuasa2004}.  Several types of writing
schemes have been developed. The first commercial MRAM employed the
field switching\cite{Savtchenko2001,Engel2005}. The field switching
requires high write energy, order of 100 pJ/bit \cite{ITRS2007},
because the field is generated by the current flowing through the wire
separated from the MTJ. Discovery of the spin transfer torque (STT)
switching method \cite{Slonczewski1996,Berger1996} substantially
decreased the write energy to the order of 100 fJ/bit
\cite{Cai2017}. However, it is still two orders of magnitude larger
than the write energy of static random-access memory, $\sim$ 1
fJ/bit. In STT switching the main contribution to the write energy is
Ohmic dissipation, i.e. Joule heating. In order to decrease the write
energy further much effort has been devoted to decreasing the critical
current density for STT-switching\cite{Yen2008,Bosu2016,Suess2017}.

Voltage-torque (VT) switching is another attractive method for low
power writing, which is based on voltage control of magnetic
anisotropy (VCMA) in a thin ferromagnetic film
\cite{Weisheit2007,Maruyama2009,Nozaki2010,Shiota2011,Nozaki2014,Lin2014,Amiri2015,Kanai2016,Grezes2016,Munira2016,Nozaki2016,Shiota2016,Nozaki2017,Cai2017,Song2017,Yamamoto2018,Ikeura2018,Matsumoto2018,Pavan2019,Yamamoto2019,Matsumoto2019}.
The mechanism of the VCMA in a MgO-based MTJ is considered as the
combination of the selective electron/hole doping into the d-electron
orbitals and the induction of a magnetic dipole moment, which affect
the electron spin through the spin-orbit
interaction\cite{Duan2008,Nakamura2009,Tsujikawa2009,Niranjan2010,Miwa2017}.
Very recently VT switching with very small write energy of about 6
fJ/bit was demonstrated by Grezes\cite{Grezes2016} et al. and
independently by Kanai et al.\cite{Kanai2016}.

The basic structure of the MTJ for the VT-MRAM is the same as that for
the STT-MRAM except for the value of the resistance area (RA)
product. The VT-MRAM has much larger RA product than that for the
STT-MRAM to suppress the current density, or energy loss by Joule
heating, at the operating voltage. Although the Joule heating at the
operating voltage reduces as the resistance increases, the read time
increases with increase of the resistance because it is determined by
the $RC$ time constant, where $R$ and $C$ are the resistance and
capacitance of the MRAM cell, respectively.  The resistance of the
MRAM cell should be designed to balance the energy consumption and
read time.

The write error rate (WER) is another key metric to characterize the
performance of the MRAM cell
\cite{Worledge2010,Min2010,Nowak2011,Sun2012,Apalkov2016,Shiota2016,Shiota2017,Ikeura2018,Yamamoto2019}
The WER of STT-MRAM can be lowered by increasing the applied current
density\cite{Worledge2010,Min2010,Nowak2011,Sun2012,Apalkov2016}. Nowak
et al. reported the WERs below 10$^{-11}$ are reported for a 4-kb
STT-MRAM chip\cite{Nowak2011}.  The WER of the VT-MRAM is still higher
than that of the STT-MRAM, which ranges form 10$^{-3}$ to
10$^{-5}$\cite{Shiota2016,Shiota2017,Yamamoto2019}.  People are
working to reduce the WER by improving materials\cite{Shiota2017} as
well as by shaping voltage pulse\cite{Ikeura2018,Yamamoto2019}.  Until
now the WER of the VT-MRAM has been studied in the high resistance
condition to eliminate the effects of STT.  However, for practical
application, the resistance should be lowered to decrease the read
time. It is important to know the minimum current density below which
the impact of STT on the WER is negligible.

In this paper, WER of a perpendicularly magnetized VT-MRAM is
theoretically studied with special attention to the impact of STT. It
is shown that the STT assists or suppresses the precessional motion of
the magnetization depending on the direction of the initial state,
i.e. up-polarized or down-polarized. There exists a characteristic
value of the current density above which the precessional motion, and
therefore magnetization switching, is forbidden by the STT for one
direction of the switching. It is found that for typical material
parameters the WER is insensitive to the STT below the current density
of $10^{10}$ A/m$^{2}$.

\begin{figure}[t]
  \centerline{ \includegraphics[width=\columnwidth]{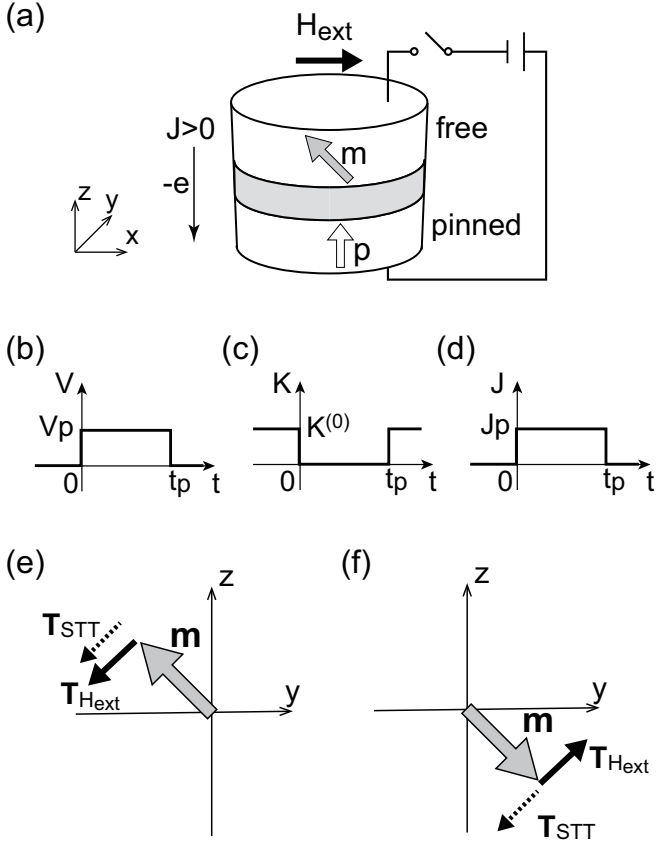}}
  \caption{\label{fig:schem} (Color online) (a) Magnetic tunnel
    junction with circular cylinder shape and definitions of Cartesian
    coordinates $(x, y, z)$.  The $x$ axis is parallel to the
    direction of external in-plane magnetic field, ${\bm H}_{\rm
      ext}$. The unit vectors ${\bm m}$ ($\bm{p}$) represents the
    direction of the magnetization in the free (reference) layer.  The
    positive current density, $J>0$, is defined as electrons flowing
    from the free layer to the reference layer.  (b) The shape of the
    voltage pulse. The duration of the pulse is represented by $t_{\rm
      p}$. The voltage during the pulse is denoted by $V_{\rm p}$.
    (c) The time dependence of the anisotropy constant $K$. The
    anisotropy constant without application of the voltage is
    represented by $K^{(0)}$.  (d) The time dependence of the current
    density $J$.  The current density during the pulse is denoted by
    $J_{\rm p}$.  (e) Directions of the torques $\bm{T}_{H_{\rm ext}}$
    and $\bm{T}_{\rm STT}$ exerted on $\bm{m}$ in $yz$-plane in the
    case that $m_{y}<0$.  (f) The same plot for $m_{y}>0$.  }
\end{figure}

\section{THEORETICAL MODEL}
The circular-shaped MTJ-nanopillar we consider is schematically shown
in Fig. \ref{fig:schem}(a).  The insulating layer is sandwiched by the
two ferromagnetic layers: the free layer and the reference layer. The
direction of the magnetization in the free layer is represented by the
unit vector $\bm{m} = (m_{x}, m_{y}, m_{z})$. The magnetization unit
vector in the reference layer $\bm{p}$ is fixed to align in the
positive $z$ direction, i.e. $\bm{p}=(0,0,1)$. The $x$ and $y$ axes
are taken to be the in-plane directions, while the $z$ axis is taken
to be the out-of-plane direction. The static external-field,
$\bm{H}_{\rm ext}$, is applied in the positive $x$ direction. The
positive current density, $J>0$, is defined as electrons flowing from
the free layer to the reference layer.  The size of the nanopillar is
assumed to be so small that the magnetization dynamics can be
described by the macrospin model.

The shape of the pulse of voltage, $V$, we assumed is shown in
Fig. \ref{fig:schem}(b), where $V_{\rm p}$ and $t_{\rm p}$ represent
the amplitude and the duration of the pulse, respectively. Without
application of the voltage the free layer is assumed to have the
out-of-plane uniaxial anisotropy which is characterized by the
anisotropy constant $K_{\rm u}^{(0)}$. Here $K_{\rm u}^{(0)}$
represents the total anisotropy comprising the crystalline anisotropy,
the interfacial anisotropy, and the shape anisotropy.  As shown in
Fig. \ref{fig:schem}(c) the anisotropy constant is assumed to decrease
to zero by application of the voltage of $V_{\rm p}$ through the VCMA
effect. During the voltage pulse the current with density of $J_{\rm
  p}$ flows through the MTJ-nanopillar as shown in
Fig. \ref{fig:schem}(d).  The time dependence of the voltage,
anisotropy constant, and current density are summarized as
\begin{equation}
  [V,K_{\rm u}, J](t) =
  \begin{cases}
    [V_{\rm p},0, J_{\rm p}] & 0\le t \le t_{\rm p}\\ [0,K_{\rm
        u}^{(0)},0] & {\rm otherwise}
  \end{cases}.
\end{equation}

The dynamics of the magnetization unit vector in the free layer is
obtained by solving the Landau Lifshitz Gilbert (LLG) equation
\begin{equation}
  \label{eq:llg}
  \frac{d\bm{m}}{dt} = -\gamma \bm{m}\times \bm{H}_{\rm eff} -\gamma
  \chi \bm{m}\times\left(\bm{m}\times\bm{p}\right) +\alpha
  \bm{m}\times\frac{d\bm{m}}{dt},
\end{equation}
where the first, second, and third terms on the right hand side
represent the torque due to the effective field $\bm{H}_{\rm eff}$,
STT, and damping torque, respectively.  The effective field comprises
the external field, anisotropy field, $\bm{H}_{\rm anis}$, and thermal
agitation field, $\bm{H}_{\rm therm}$, as
\begin{equation}
  \bm{H}_{\rm eff} = \bm{H}_{\rm ext} + \bm{H}_{\rm anis} +
  \bm{H}_{\rm therm}.
\end{equation}
The anisotropy field is defined as
\begin{equation}
  \bm{H}_{\rm anis} = \frac{2 K_{u}(t)}{\mu_{0}M_{\rm
      s}}m_{z}(t)\bm{e}_{z},
\end{equation}
where $\bm{e}_{z}$ is the unit vector in the positive $z$ direction.
The thermal agitation field is determined by the
fluctuation-dissipation theorem
\cite{Brown1963,Callen1951,Callen1952a,Callen1952b,Greene1952} and
satisfies the following relations
\begin{align}
  & \left\langle H_{\rm therm}^{i}(t) \right\rangle = 0, \\ &
  \left\langle H_{\rm therm}^{i}(t)\, H_{\rm
    therm}^{j}(t')\ \right\rangle = \mu\,\delta_{i,j}\, \delta(t-t'),
\end{align}
where indices $i$, $j$ denote the $x$, $y$, and $z$ components of the
thermal agitation field. $\delta_{i,j}$ represents Kronecker's delta,
and $\delta(t-t')$ represents Dirac's delta function
The coefficient $\mu$ is given by
\begin{equation}
  \mu = \frac{2\alpha k_{\rm B} T}{\gamma\, \mu_{0}\, M_{\rm s}\,
    \Omega},
\end{equation}
where $k_{\rm B}$ is the Boltzmann constant, $T$ is temperature, and
$\Omega$ is the volume of the free layer.  The coefficient of the STT
is defined as
\begin{equation}
  \chi = \frac{\hbar P J(t)}{2 e \mu_{0} M_{\rm s} d},
\end{equation}
where $P$ is the spin polarization of the current, $e$ is the
elementary charge, $d$ is the thickness of the free
layer\cite{Slonczewski1996,Stiles2005}. Here the angle dependence of
$\chi$ is neglected for simplicity.

The following parameters are assumed for numerical calculations:
$\alpha$ = 0.1, $K_{\rm u}$ = 0.11 MJ/m$^{3}$, $M_{\rm s}$ = 0.955
MA/m \cite{Yamamoto2019}.  The magnitude of the external-field is
$H_{\rm ext}$ = 970 Oe.  The diameter of the free layer is 40 nm. The
thickness of the free layer is $d$= 1.1 nm. The spin polarization of
current is $P=0.6$. The WERs are calculated from 10$^{6}$ trials.

\section{RESULTS AND DISCUSSION}
Before showing the numerical results let us discuss the role of STT on
the dynamics of $\bm{m}$. Since $\bm{H}_{\rm ext}$ is the static
external-field, the torque due to $\bm{H}_{\rm ext}$ is exerted on
$\bm{m}$ all the time. The STT exists only during the pulse, where the
anisotropy constant is zero. During the pulse the following two kinds
of torques give the dominant contributions to the magnetization
dynamics: One is the external field torque,
\begin{equation}
  \bm{T}_{H_{\rm ext}} = -\gamma \bm{m}\times\bm{H}_{\rm ext},
\end{equation}
and the other is the STT,
\begin{equation}
  \bm{T}_{\rm STT} = -\gamma \chi
  \bm{m}\times\left(\bm{m}\times\bm{p}\right).
\end{equation}
Neglecting the thermal agitation and damping, the trajectory of
magnetization precession for switching can be well approximated by the
the semi-arc on the $yz$ plane. Therefore the vector
$\bm{m}\times\bm{p}$ is parallel or anti-parallel to the
external-field depending on the sign of $m_{y}$.

For the switching from the up-state ($m_{z}>0$) to the down-state
($m_{z}<0$), the vector $\bm{m}\times\bm{p}$ is parallel to the
external-field, and therefore $\bm{T}_{\rm STT}$ is parallel to
$\bm{T}_{H_{\rm ext}}$ as shown in Fig. \ref{fig:schem}(e). The
angular velocity of the precessional motion of $\bm{m}$ is increased
by the STT as if the external-field is enhanced.

On the contrary, for the switching from the down-state to the
up-state, $\bm{T}_{\rm STT}$ is anti-parallel to $\bm{T}_{H_{\rm
    ext}}$ as shown in Fig. \ref{fig:schem}(f). The angular velocity
of $\bm{m}$ is decreased by the STT as if the external field is
reduced. There exists a characteristic current density above which the
STT overcomes the external field torque, which is obtained by solving
$T_{H_{\rm ext}} = T_{\rm STT}$ as
\begin{equation}
  \label{eq:jcirt}
  J_{\rm p}^{(\rm c)} = \frac{2e\mu_{0}M_{\rm s}d}{\hbar P}H_{\rm ext}.
\end{equation}
For the parameters stated before the value of the characteristic
current density is $J_{\rm p}^{(\rm c)} = 5.4\times 10^{11}$
A/m$^{2}$, which is as large as the typical value of the critical
current density for the STT switching.

\begin{figure}[t]
  \centerline{ \includegraphics[width=\columnwidth]{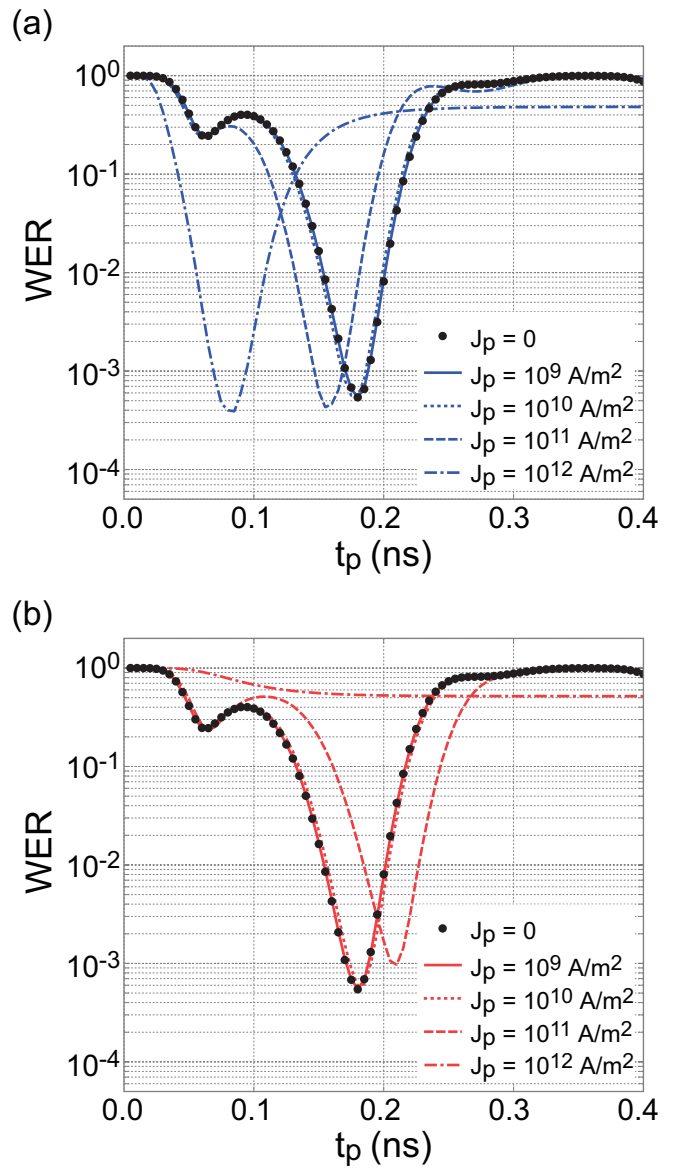}}
  \caption{(Color online) (a) Pulse width ($t_{\rm p}$) dependence of
    the WER for the switching from the up-state to the down-state. The
    results in the absence of current, i.e $J_{\rm p}$ =0 are
    represented by the circles. The results for the current density of
    $J_{\rm p} = 10^{9}$, $10^{10}$, $10^{11}$, $10^{12}$ A/m$^{2}$ are
    represented by the solid, dotted, dashed, and dot-dashed curves.
    (b) The same plot for switching from down-state to the up-state.
    \label{fig:tp_WER}
  }
\end{figure}

The value of $J_{\rm p}^{(\rm c)}$ gives a rough estimation of the
current density above which the switching from the down-state to the
up-state is forbidden.  Even if the current density is smaller than
$J_{\rm p}^{(\rm c)}$ the STT can affect the precessional motion and
increase or decrease the WER. For quantitative understanding of the
impact of STT on the WER we perform numerical simulations based on
Eq. \eqref{eq:llg}.  There are two approaches to obtain the WER
starting from Eq. \eqref{eq:llg}.  One is the Fokker-Planck-equation
approach\cite{Brown1963,Apalkov2005b,Apalkov2005b,Tzoufras2018} and
the other is the Langevin-equation
approach\cite{Shiota2016,Ikeura2018,Yamamoto2018,Matsumoto2018,Yamamoto2019,Matsumoto2019}.
In principle these two approaches give the same results because they
are based on the same LLG equation. Here we employ the
Langevin-equation approach because we have many experiences on this
approach and have reproduced the experimentally observed WER very well
as reported in Refs. \citenum{Yamamoto2018} and
\citenum{Yamamoto2019}.

Figure \ref{fig:tp_WER}(a) shows the $t_{\rm p}$ dependence of the WER
for the switching from the up-state to the down-state. The initial
states are prepared by relaxing the magnetization from the equilibrium
direction at $T=0$ with $m_{z}>0$ for 5 ns before the beginning of the
pulse.  The success or failure of switching is determined by the sign
of $m_{z}$ at 5 ns after the end of the pulse.  During this 5 ns the
voltage is not applied, and therefore the magnetization relaxes to the
equilibrium directions.  The distributions of $m_{z}$ at the beginning
of the pulse and at 5 ns after the end of the pulse are well localized
around the equilibrium directions (see APPENDIX A).  The results for
$J_{\rm p}$ =0 are represented by the circles.  Since the WER
satisfies the binomial distribution, the standard deviation of the WER
is given by $\sqrt{ q (1-q) / N}$, where $q$ is the switching
probability, $1-p$ is the WER, and $N$ is the number of trials.  For
$J=0$ the WER takes the minimum value of 5.46$\times$10$^{-4}$ at
$t_{\rm p} = 0.18$ ns. The corresponding standard deviation for
$N=10^6$ trials is 2.34$\times$10$^{-5}$, which is smaller than the
radius of circles plotted in Figs. \ref{fig:tp_WER}(a) and
\ref{fig:tp_WER}(b).  The results for the current density of $J_{\rm
  p} = 10^{9}$, $10^{10}$, $10^{11}$, $10^{12}$ are represented by the
solid, dotted, dashed, and dot-dashed curves.  Below $10^{10}$
A/m$^{2}$ the $t_{\rm p}$ dependence of WER is almost the same as that
for $J_{\rm p}$ =0 because $T_{\rm STT}$ is much smaller than
$T_{H_{\rm ext}}$.  Above $10^{11}$ A/m$^{2}$ the pulse width at which
the WER is minimized decreases with increase of the current density,
because the STT assists the precessional motion around the
external-field.

Figure \ref{fig:tp_WER}(b) shows the $t_{\rm p}$ dependence of the WER
for the switching from the down-state to the up-state. Similar to
Fig. \ref{fig:tp_WER}(a) the $t_{\rm p}$ dependence of the WER is
almost the same as that for $J_{\rm p}$ =0 for $J_{\rm p} \leq
10^{10}$ A/m$^{2}$.  At $J_{\rm p} = 10^{11}$ A/m$^{2}$ the pulse
width at which the WER is minimized increases with increase of the
current density, because the STT suppresses the precessional motion
around the external-field. At $J_{\rm p} = 10^{12}$ A/m$^{2}$ the dip
in the $t_{\rm p}$ dependence of WER disappears as shown by the
dot-dashed curve because $T_{\rm STT}$ exceeds $T_{H_{\rm ext}}$ much
earlier than one half of the precession period.  The magnitude of the
STT is proportional to the sine of the relative angle, $\theta_{\rm
  r}$, between $\bm{m}$ and $\bm{p}$. The relative angle is
$\theta_{\rm r}=\pi$ at the initial down-state and decreases as the
magnetization precesses toward the up-sate ($\theta_{\rm r}=0$).  The
precession stops once $\theta_{\rm r}$ reaches a certain critical
angle where the external-field torque is canceled with the STT. For
the switching from the down-state to the up-state the critical angle
increases with increase of the current density.

\begin{figure}[t]
  \centerline{ \includegraphics[width=\columnwidth]{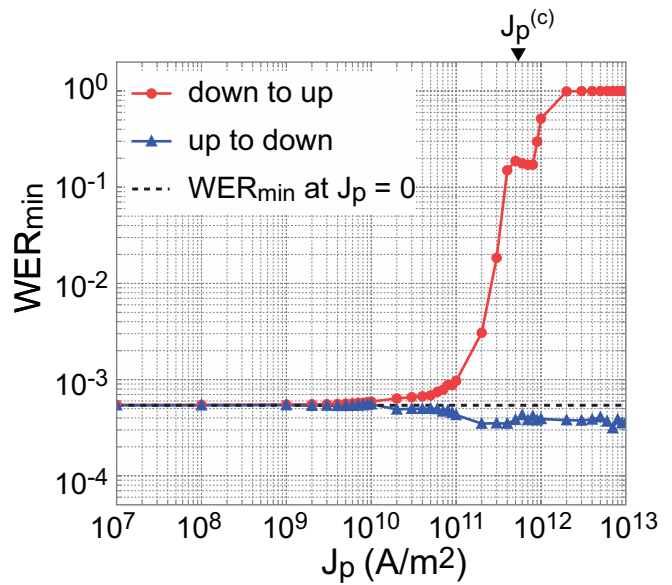}}
  \caption{(Color online) Current density ($J_{\rm p}$) dependence of
    WER$_{\rm min}$. The results for switching from down-state to the
    up-state are shown by the blue triangles. Those for switching from
    up-state to the down-state. are shown by the red circles. The value
    of WER$_{\rm min}$ at $J_{\rm p}=0$ is indicated by the dotted
    line. The value of the characteristic current density, $J_{\rm
      p}^{(c)}$, is indicated by the black triangle on the top of the
    frame.
    \label{fig:Jp_WER}
  }
\end{figure}

In Fig. \ref{fig:Jp_WER} the minimum values of WER, WER$_{\rm min}$,
are plotted as a function of the current density. The results for the
switching from the up-state to the down-state are shown by the blue
triangles. The value of WER$_{\rm min}$ at $J_{\rm p}=0$ is indicated
by the dotted line as a guide. Below the current density of $10^{11}$
A/m$^{2}$ the WER$_{\rm min}$ takes almost the same value as that at
$J_{\rm p}=0$. It shows a shallow decrease above $10^{11}$ A/m$^{2}$.
The red circles show WER$_{\rm min}$ for the switching from the
down-state to the up-state. One can easily confirm that below the
current density of $10^{10}$ A/m$^{2}$ the WER$_{\rm min}$ takes
almost the same value as that at $J_{\rm p}=0$. It shows a shallow
increase until $J_{\rm p}$ reaches $10^{11}$ A/m$^{2}$. Above the
current density of $10^{11}$ A/m$^{2}$ it shows a rapid increase and
reaches almost unity at $J_{\rm p}$ = $2\times 10^{12}$ A/m$^{2}$. At
around $J_{\rm p} = J_{\rm p}^{(\rm c)}$ there appears a plateau where
the WER is insensitive to the variation of $J_{\rm p}$.

\begin{figure}[t]
  \centerline{ \includegraphics[width=0.9\columnwidth]{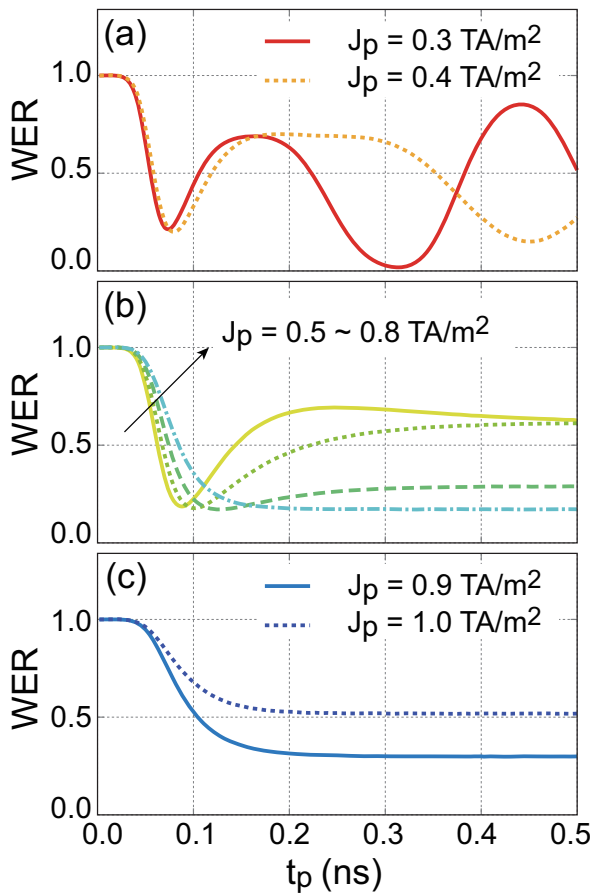}}
  \caption{(Color online) (a) Pulse width ($t_{\rm p}$) dependence of
    WER at $J_{\rm p}$=0.3 and 0.4 TA/m$^{2}$ (b) The same plot at
    $J_{\rm p}$=0.5 $\sim$ 0.8 TA/m$^{2}$.  (c) The same plot at $J_{\rm
      p}$=0.9 and 1.0 TA/m$^{2}$.
    \label{fig:pl}
  }
\end{figure}

In order to understand the mechanism for appearance of the plateau,
let us look at the $t_{\rm p}$-dependence of the WER at the current
density around $J_{\rm p}^{(\rm c)}$. Figure \ref{fig:pl}(a) shows the
$t_{\rm p}$-dependence of WER at $J_{\rm p}$=0.3 and 0.4 TA/m$^{2}$,
where the WER$_{\rm min}$ shows a rapid increase. The WER takes the
minimum value at the second dip which corresponds to one half of the
precession period. The appearance of the first dip, or the appearance
of the bump between the first and the second dips, is originated from
the thermally induced precession-orbit transition of magnetization as
discussed in Ref. \citenum{Yamamoto2018}. The position of the bump
corresponds to one quarter of the precession period, at which the
magnetization is on the equator plane on the Bloch sphere,
i.e. $m_{z}=0$. Since the magnetization around this direction has high
anisotropy energy in the relaxation process it takes long time for the
magnetization to relax to the equilibrium direction, the up-state or
the down-state. Therefore the probability of switch failure or the WER
is enhanced around the pulse width of one quarter of the precession
period.  As the current density increases from $J_{\rm p}$=0.3 to 0.4
TA/m$^{2}$ the position of the second dip moves to the longer $t_{\rm
  p}$ and the minimum value increases.

Further increase of current density eliminates the second dip and
moves the position of the WER$_{\rm min}$ to the longer $t_{\rm p}$ as
shown in Fig. \ref{fig:pl}(b). From $J_{\rm p}$=0.5 to 0.7 TA/m$^{2}$
the increase of the current density does not change the value of the
WER$_{\rm min}$ very much but decreases WER at $t_{\rm p}$ longer than
the first dip because in this range of the current density the STT
exceeds external-field torque around one quarter of the precession
period.  At $J_{\rm p}$= 0.8 TA/m$^{2}$ the first dip, or the bump,
disappears. Above the current density of 0.9 TA/m$^{2}$ the WER$_{\rm
  min}$ increases with increase of $J_{\rm p}$ as shown in
Fig. \ref{fig:pl}(c).

\section{SUMMARY}
In summary the impact of STT on the WER of a VT-MRAM is theoretically
investigated. The characteristic value of the current density above
which the precessional motion is forbidden by the STT is derived by
balancing the STT and the external-field torque.  The WER is
insensitive to the STT at the current density below $10^{10}$
A/m$^{2}$.

\begin{acknowledgments}
  This work was partly supported by JSPS KAKENHI Grant Number 19H01108,
  and the ImPACT Program of the Council for Science, Technology and
  Innovation.
\end{acknowledgments}

\appendix

\section{Distribution of $m_{z}$}

\begin{figure}[t]
  \centerline{ \includegraphics[width=\columnwidth]{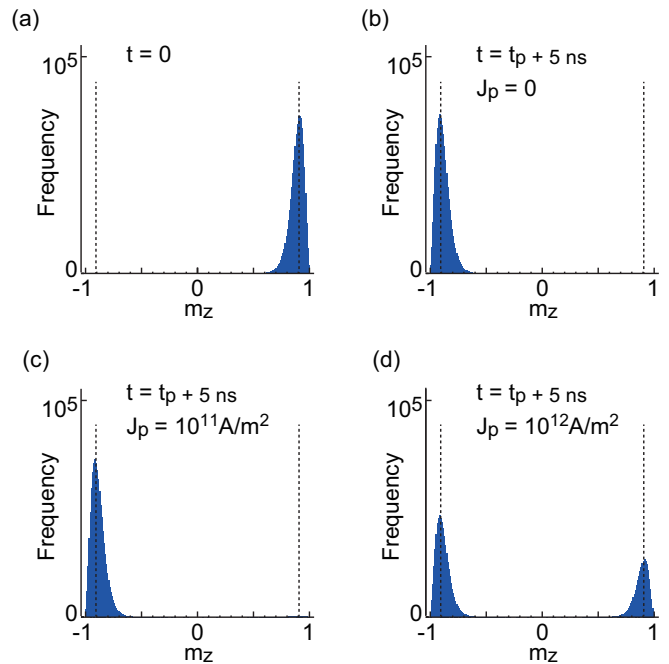}}
  \caption{
    (Color online)
    Distribution of $m_{z}$ for the switching from the up-state to the
    down-state.
    (a) Distribution of $m_{z}$ at the
    beginning of pulse ($t=0$), which is independent of the value of
    $J_{p}$.
    (b) Distribution of $m_{z}$ at 5 ns after the end of
    pulse ($t=t_{\rm p}$ + 5 ns) for the current density of $J_{p}=0$.
    The pulse width is set $t_{\rm p}$ = 0.18 ns, at which the WER is
    minimized.
    (c) The same plot as (b) for $J_{\rm p} = 10^{11}$ A/m$^{2}$.
    (d) The same plot as (b) for $J_{\rm p} = 10^{12}$ A/m$^{2}$.
    In all panels the vertical dotted lines represent the values of
    $m_{z}$ of the equilibrium directions at $T=0$.
    \label{fig:hist_ud}
  }
\end{figure}

In this section we discuss the distributions of $m_{z}$ at the beginning
of pulse and at the 5 ns after the end of pulse. The
states at the beginning of the pulse are prepared by relaxing the
magnetization from the equilibrium direction at $T=0$ for 5 ns. The
success or failure of switching is determined by the sign of $m_{z}$
at 5 ns after the end of the pulse.
The relaxation time of 5 ns is set to be long enough for magnetization
to be relaxed around the equilibrium directions. To confirm the
validity of this procedure we show the distributions of $m_{z}$
for the switching from the up-state to the down-state as histograms in
Figs. \ref{fig:hist_ud}(a) --  \ref{fig:hist_ud}(d).

The distribution of $m_{z}$ at the beginning of pulse ($t=0$) is shown in
Fig. \ref{fig:hist_ud}(a) where the values of $m_{z}$ of the
equilibrium directions at $T=0$ are indicated by the vertical dotted
lines. The initial
distribution is independent of the value of $J_{\rm p}$ because it is
prepared by relaxing the magnetization without applying current. The
distribution is well localized in the vicinity of the positive
equilibrium value.

Application of the voltage pulse induces the precessional motion of
magnetization and switches the magnetization direction with a certain
probability. Then the magnetization relaxes to the equilibrium
directions because $J=0$ and $K_{\rm u} = K_{\rm u}^{(0)}$ for
$t \ge t_{\rm p}$. The success or failure of switching is determined
by the sign of $m_{z}$ at $t=t_{\rm p}$ + 5 ns.
In Figs. 5(b), (c), and (d) the distributions at $t=t_{\rm p}$ + 5 ns
are plotted for $J_{\rm p}$ = 0, $10^{11}$, and $10^{12}$ A/m$^{2}$ ,
respectively. In these figures the pulse width is assumed to be
$t_{\rm p}=0.18$ ns at which the WER for $J_{\rm p}$ = 0 is minimized.
As shown in Figs. 5 (b) -- (d) the distributions are well localized in the
vicinity of the equilibrium values, which enables to clearly determine the
success or failure of switching.


\begin{thebibliography}{52}%
\makeatletter
\providecommand \@ifxundefined [1]{%
 \@ifx{#1\undefined}
}%
\providecommand \@ifnum [1]{%
 \ifnum #1\expandafter \@firstoftwo
 \else \expandafter \@secondoftwo
 \fi
}%
\providecommand \@ifx [1]{%
 \ifx #1\expandafter \@firstoftwo
 \else \expandafter \@secondoftwo
 \fi
}%
\providecommand \natexlab [1]{#1}%
\providecommand \enquote  [1]{``#1''}%
\providecommand \bibnamefont  [1]{#1}%
\providecommand \bibfnamefont [1]{#1}%
\providecommand \citenamefont [1]{#1}%
\providecommand \href@noop [0]{\@secondoftwo}%
\providecommand \href [0]{\begingroup \@sanitize@url \@href}%
\providecommand \@href[1]{\@@startlink{#1}\@@href}%
\providecommand \@@href[1]{\endgroup#1\@@endlink}%
\providecommand \@sanitize@url [0]{\catcode `\\12\catcode `\$12\catcode
  `\&12\catcode `\#12\catcode `\^12\catcode `\_12\catcode `\%12\relax}%
\providecommand \@@startlink[1]{}%
\providecommand \@@endlink[0]{}%
\providecommand \url  [0]{\begingroup\@sanitize@url \@url }%
\providecommand \@url [1]{\endgroup\@href {#1}{\urlprefix }}%
\providecommand \urlprefix  [0]{URL }%
\providecommand \Eprint [0]{\href }%
\providecommand \doibase [0]{http://dx.doi.org/}%
\providecommand \selectlanguage [0]{\@gobble}%
\providecommand \bibinfo  [0]{\@secondoftwo}%
\providecommand \bibfield  [0]{\@secondoftwo}%
\providecommand \translation [1]{[#1]}%
\providecommand \BibitemOpen [0]{}%
\providecommand \bibitemStop [0]{}%
\providecommand \bibitemNoStop [0]{.\EOS\space}%
\providecommand \EOS [0]{\spacefactor3000\relax}%
\providecommand \BibitemShut  [1]{\csname bibitem#1\endcsname}%
\let\auto@bib@innerbib\@empty
\bibitem [{\citenamefont {Yuasa}\ \emph {et~al.}(2013)\citenamefont {Yuasa},
  \citenamefont {Fukushima}, \citenamefont {Yakushiji}, \citenamefont {Nozaki},
  \citenamefont {Konoto}, \citenamefont {Maehara}, \citenamefont {Kubota},
  \citenamefont {Taniguchi}, \citenamefont {Arai}, \citenamefont {Imamura},
  \citenamefont {Ando}, \citenamefont {Shiota}, \citenamefont {Bonell},
  \citenamefont {Suzuki}, \citenamefont {Shimomura}, \citenamefont {Kitagawa},
  \citenamefont {Ito}, \citenamefont {Fujita}, \citenamefont {Abe},
  \citenamefont {Nomura}, \citenamefont {Noguchi},\ and\ \citenamefont
  {Yoda}}]{Yuasa2013}%
  \BibitemOpen
  \bibfield  {author} {\bibinfo {author} {\bibfnamefont {S.}~\bibnamefont
  {Yuasa}}, \bibinfo {author} {\bibfnamefont {A.}~\bibnamefont {Fukushima}},
  \bibinfo {author} {\bibfnamefont {K.}~\bibnamefont {Yakushiji}}, \bibinfo
  {author} {\bibfnamefont {T.}~\bibnamefont {Nozaki}}, \bibinfo {author}
  {\bibfnamefont {M.}~\bibnamefont {Konoto}}, \bibinfo {author} {\bibfnamefont
  {H.}~\bibnamefont {Maehara}}, \bibinfo {author} {\bibfnamefont
  {H.}~\bibnamefont {Kubota}}, \bibinfo {author} {\bibfnamefont
  {T.}~\bibnamefont {Taniguchi}}, \bibinfo {author} {\bibfnamefont
  {H.}~\bibnamefont {Arai}}, \bibinfo {author} {\bibfnamefont {H.}~\bibnamefont
  {Imamura}}, \bibinfo {author} {\bibfnamefont {K.}~\bibnamefont {Ando}},
  \bibinfo {author} {\bibfnamefont {Y.}~\bibnamefont {Shiota}}, \bibinfo
  {author} {\bibfnamefont {F.}~\bibnamefont {Bonell}}, \bibinfo {author}
  {\bibfnamefont {Y.}~\bibnamefont {Suzuki}}, \bibinfo {author} {\bibfnamefont
  {N.}~\bibnamefont {Shimomura}}, \bibinfo {author} {\bibfnamefont
  {E.}~\bibnamefont {Kitagawa}}, \bibinfo {author} {\bibfnamefont
  {J.}~\bibnamefont {Ito}}, \bibinfo {author} {\bibfnamefont {S.}~\bibnamefont
  {Fujita}}, \bibinfo {author} {\bibfnamefont {K.}~\bibnamefont {Abe}},
  \bibinfo {author} {\bibfnamefont {K.}~\bibnamefont {Nomura}}, \bibinfo
  {author} {\bibfnamefont {H.}~\bibnamefont {Noguchi}}, \ and\ \bibinfo
  {author} {\bibfnamefont {H.}~\bibnamefont {Yoda}},\ }\bibfield  {title}
  {\enquote {\bibinfo {title} {{Future prospects of MRAM technologies}},}\ }in\
  \href {\doibase 10.1109/IEDM.2013.6724549} {\emph {\bibinfo {booktitle}
  {Technical Digest - International Electron Devices Meeting, IEDM}}}\
  (\bibinfo {year} {2013})\BibitemShut {NoStop}%
\bibitem [{\citenamefont {Apalkov}\ \emph {et~al.}(2016)\citenamefont
  {Apalkov}, \citenamefont {Dieny},\ and\ \citenamefont
  {Slaughter}}]{Apalkov2016}%
  \BibitemOpen
  \bibfield  {author} {\bibinfo {author} {\bibfnamefont {Dmytro}\ \bibnamefont
  {Apalkov}}, \bibinfo {author} {\bibfnamefont {Bernard}\ \bibnamefont
  {Dieny}}, \ and\ \bibinfo {author} {\bibfnamefont {J.~M.}\ \bibnamefont
  {Slaughter}},\ }\bibfield  {title} {\enquote {\bibinfo {title}
  {{Magnetoresistive Random Access Memory}},}\ }\href {\doibase
  10.1109/JPROC.2016.2590142} {\bibfield  {journal} {\bibinfo  {journal}
  {Proceedings of the IEEE}\ }\textbf {\bibinfo {volume} {104}},\ \bibinfo
  {pages} {1796} (\bibinfo {year} {2016})}\BibitemShut {NoStop}%
\bibitem [{\citenamefont {Sbiaa}\ and\ \citenamefont
  {Piramanayagam}(2017)}]{Sbiaa2017}%
  \BibitemOpen
  \bibfield  {author} {\bibinfo {author} {\bibfnamefont {Rachid}\ \bibnamefont
  {Sbiaa}}\ and\ \bibinfo {author} {\bibfnamefont {S~N}\ \bibnamefont
  {Piramanayagam}},\ }\bibfield  {title} {\enquote {\bibinfo {title} {{Recent
  Developments in Spin Transfer Torque MRAM}},}\ }\href {\doibase
  10.1002/pssr.201700163} {\bibfield  {journal} {\bibinfo  {journal} {physica
  status solidi (RRL) - Rapid Research Letters}\ }\textbf {\bibinfo {volume}
  {11}},\ \bibinfo {pages} {1700163} (\bibinfo {year} {2017})}\BibitemShut
  {NoStop}%
\bibitem [{\citenamefont {Cai}\ \emph {et~al.}(2017)\citenamefont {Cai},
  \citenamefont {Kang}, \citenamefont {Wang}, \citenamefont {Naviner},
  \citenamefont {Yang},\ and\ \citenamefont {Zhao}}]{Cai2017}%
  \BibitemOpen
  \bibfield  {author} {\bibinfo {author} {\bibfnamefont {Hao}\ \bibnamefont
  {Cai}}, \bibinfo {author} {\bibfnamefont {Wang}\ \bibnamefont {Kang}},
  \bibinfo {author} {\bibfnamefont {You}\ \bibnamefont {Wang}}, \bibinfo
  {author} {\bibfnamefont {Lirida}\ \bibnamefont {Naviner}}, \bibinfo {author}
  {\bibfnamefont {Jun}\ \bibnamefont {Yang}}, \ and\ \bibinfo {author}
  {\bibfnamefont {Weisheng}\ \bibnamefont {Zhao}},\ }\bibfield  {title}
  {\enquote {\bibinfo {title} {{High Performance MRAM with Spin-Transfer-Torque
  and Voltage-Controlled Magnetic Anisotropy Effects}},}\ }\href {\doibase
  10.3390/app7090929} {\bibfield  {journal} {\bibinfo  {journal} {Applied
  Sciences}\ }\textbf {\bibinfo {volume} {7}},\ \bibinfo {pages} {929}
  (\bibinfo {year} {2017})}\BibitemShut {NoStop}%
\bibitem [{\citenamefont {Parkin}\ \emph {et~al.}(2004)\citenamefont {Parkin},
  \citenamefont {Kaiser}, \citenamefont {Panchula}, \citenamefont {Rice},
  \citenamefont {Hughes}, \citenamefont {Samant},\ and\ \citenamefont
  {Yang}}]{Parkin2004}%
  \BibitemOpen
  \bibfield  {author} {\bibinfo {author} {\bibfnamefont {Stuart S.~P.}\
  \bibnamefont {Parkin}}, \bibinfo {author} {\bibfnamefont {Christian}\
  \bibnamefont {Kaiser}}, \bibinfo {author} {\bibfnamefont {Alex}\ \bibnamefont
  {Panchula}}, \bibinfo {author} {\bibfnamefont {Philip~M.}\ \bibnamefont
  {Rice}}, \bibinfo {author} {\bibfnamefont {Brian}\ \bibnamefont {Hughes}},
  \bibinfo {author} {\bibfnamefont {Mahesh}\ \bibnamefont {Samant}}, \ and\
  \bibinfo {author} {\bibfnamefont {See-Hun}\ \bibnamefont {Yang}},\ }\bibfield
   {title} {\enquote {\bibinfo {title} {{Giant tunnelling magnetoresistance at
  room temperature with MgO (100) tunnel barriers}},}\ }\href {\doibase
  10.1038/nmat1256} {\bibfield  {journal} {\bibinfo  {journal} {Nature
  Materials}\ }\textbf {\bibinfo {volume} {3}},\ \bibinfo {pages} {862}
  (\bibinfo {year} {2004})}\BibitemShut {NoStop}%
\bibitem [{\citenamefont {Yuasa}\ \emph {et~al.}(2004)\citenamefont {Yuasa},
  \citenamefont {Nagahama}, \citenamefont {Fukushima}, \citenamefont {Suzuki},\
  and\ \citenamefont {Ando}}]{yuasa2004}%
  \BibitemOpen
  \bibfield  {author} {\bibinfo {author} {\bibfnamefont {Shinji}\ \bibnamefont
  {Yuasa}}, \bibinfo {author} {\bibfnamefont {Taro}\ \bibnamefont {Nagahama}},
  \bibinfo {author} {\bibfnamefont {Akio}\ \bibnamefont {Fukushima}}, \bibinfo
  {author} {\bibfnamefont {Yoshishige}\ \bibnamefont {Suzuki}}, \ and\ \bibinfo
  {author} {\bibfnamefont {Koji}\ \bibnamefont {Ando}},\ }\bibfield  {title}
  {\enquote {\bibinfo {title} {{Giant room-temperature magnetoresistance in
  single-crystal Fe/MgO/Fe magnetic tunnel junctions}},}\ }\href {\doibase
  10.1038/nmat1257} {\bibfield  {journal} {\bibinfo  {journal} {Nature
  Materials}\ }\textbf {\bibinfo {volume} {3}},\ \bibinfo {pages} {868}
  (\bibinfo {year} {2004})}\BibitemShut {NoStop}%
\bibitem [{\citenamefont {Savtchenko}\ \emph {et~al.}(2001)\citenamefont
  {Savtchenko}, \citenamefont {Korkin}, \citenamefont {Engel}, \citenamefont
  {Rizzo}, \citenamefont {Deherrera},\ and\ \citenamefont
  {Janesky}}]{Savtchenko2001}%
  \BibitemOpen
  \bibfield  {author} {\bibinfo {author} {\bibfnamefont {Leonid}\ \bibnamefont
  {Savtchenko}}, \bibinfo {author} {\bibfnamefont {A.A.}\ \bibnamefont
  {Korkin}}, \bibinfo {author} {\bibfnamefont {B.N.}\ \bibnamefont {Engel}},
  \bibinfo {author} {\bibfnamefont {N.D.}\ \bibnamefont {Rizzo}}, \bibinfo
  {author} {\bibfnamefont {M.F.}\ \bibnamefont {Deherrera}}, \ and\ \bibinfo
  {author} {\bibfnamefont {J.A.}\ \bibnamefont {Janesky}},\ }\bibfield  {title}
  {\enquote {\bibinfo {title} {Method of writing to scalable magnetoresistance
  random access memory element},}\ }\href@noop {} {\bibfield  {journal}
  {\bibinfo  {journal} {{US} Patent 6,545,906 B1}\ } (\bibinfo {year}
  {2001})}\BibitemShut {NoStop}%
\bibitem [{\citenamefont {Engel}\ \emph {et~al.}(2005)\citenamefont {Engel},
  \citenamefont {Akerman}, \citenamefont {Butcher}, \citenamefont {Dave},
  \citenamefont {DeHerrera}, \citenamefont {Durlam}, \citenamefont
  {Grynkewich}, \citenamefont {Janesky}, \citenamefont {Pietambaram},
  \citenamefont {Rizzo}, \citenamefont {Slaughter}, \citenamefont {Smith},
  \citenamefont {Sun},\ and\ \citenamefont {Tehrani}}]{Engel2005}%
  \BibitemOpen
  \bibfield  {author} {\bibinfo {author} {\bibfnamefont {B.N.}\ \bibnamefont
  {Engel}}, \bibinfo {author} {\bibfnamefont {J.}~\bibnamefont {Akerman}},
  \bibinfo {author} {\bibfnamefont {B.}~\bibnamefont {Butcher}}, \bibinfo
  {author} {\bibfnamefont {R.W.}\ \bibnamefont {Dave}}, \bibinfo {author}
  {\bibfnamefont {M.}~\bibnamefont {DeHerrera}}, \bibinfo {author}
  {\bibfnamefont {M.}~\bibnamefont {Durlam}}, \bibinfo {author} {\bibfnamefont
  {G.}~\bibnamefont {Grynkewich}}, \bibinfo {author} {\bibfnamefont
  {J.}~\bibnamefont {Janesky}}, \bibinfo {author} {\bibfnamefont {S.V.}\
  \bibnamefont {Pietambaram}}, \bibinfo {author} {\bibfnamefont {N.D.}\
  \bibnamefont {Rizzo}}, \bibinfo {author} {\bibfnamefont {J.M.}\ \bibnamefont
  {Slaughter}}, \bibinfo {author} {\bibfnamefont {K.}~\bibnamefont {Smith}},
  \bibinfo {author} {\bibfnamefont {J.J.}\ \bibnamefont {Sun}}, \ and\ \bibinfo
  {author} {\bibfnamefont {S.}~\bibnamefont {Tehrani}},\ }\bibfield  {title}
  {\enquote {\bibinfo {title} {{A 4-Mb toggle MRAM based on a novel bit and
  switching method}},}\ }\href {\doibase 10.1109/TMAG.2004.840847} {\bibfield
  {journal} {\bibinfo  {journal} {IEEE Transactions on Magnetics}\ }\textbf
  {\bibinfo {volume} {41}},\ \bibinfo {pages} {132} (\bibinfo {year}
  {2005})}\BibitemShut {NoStop}%
\bibitem [{ITR()}]{ITRS2007}%
  \BibitemOpen
  \href@noop {} {}\bibinfo {note} {The international thechnology roadmap for
  semiconductors(ITRS): 2007, Emerging Research Devices, page 7, Table ERD3
  (2007)}\BibitemShut {NoStop}%
\bibitem [{\citenamefont {Slonczewski}(1996)}]{Slonczewski1996}%
  \BibitemOpen
  \bibfield  {author} {\bibinfo {author} {\bibfnamefont {J.C.}\ \bibnamefont
  {Slonczewski}},\ }\bibfield  {title} {\enquote {\bibinfo {title}
  {{Current-driven excitation of magnetic multilayers}},}\ }\href {\doibase
  10.1016/0304-8853(96)00062-5} {\bibfield  {journal} {\bibinfo  {journal}
  {Journal of Magnetism and Magnetic Materials}\ }\textbf {\bibinfo {volume}
  {159}},\ \bibinfo {pages} {L1} (\bibinfo {year} {1996})}\BibitemShut
  {NoStop}%
\bibitem [{\citenamefont {Berger}(1996)}]{Berger1996}%
  \BibitemOpen
  \bibfield  {author} {\bibinfo {author} {\bibfnamefont {L.}~\bibnamefont
  {Berger}},\ }\bibfield  {title} {\enquote {\bibinfo {title} {{Emission of
  spin waves by a magnetic multilayer traversed by a current}},}\ }\href
  {\doibase 10.1103/PhysRevB.54.9353} {\bibfield  {journal} {\bibinfo
  {journal} {Physical Review B}\ }\textbf {\bibinfo {volume} {54}},\ \bibinfo
  {pages} {9353} (\bibinfo {year} {1996})}\BibitemShut {NoStop}%
\bibitem [{\citenamefont {Yen}\ \emph {et~al.}(2008)\citenamefont {Yen},
  \citenamefont {Chen}, \citenamefont {Wang}, \citenamefont {Lee},
  \citenamefont {Shen}, \citenamefont {Yang}, \citenamefont {Tsai},
  \citenamefont {Hung}, \citenamefont {Shen}, \citenamefont {Tsai},\ and\
  \citenamefont {Kao}}]{Yen2008}%
  \BibitemOpen
  \bibfield  {author} {\bibinfo {author} {\bibfnamefont {Cheng-Tyng}\
  \bibnamefont {Yen}}, \bibinfo {author} {\bibfnamefont {Wei-Chuan}\
  \bibnamefont {Chen}}, \bibinfo {author} {\bibfnamefont {Ding-Yeong}\
  \bibnamefont {Wang}}, \bibinfo {author} {\bibfnamefont {Yuan-Jen}\
  \bibnamefont {Lee}}, \bibinfo {author} {\bibfnamefont {Chih-Ta}\ \bibnamefont
  {Shen}}, \bibinfo {author} {\bibfnamefont {Shan-Yi}\ \bibnamefont {Yang}},
  \bibinfo {author} {\bibfnamefont {Ching-Hsiang}\ \bibnamefont {Tsai}},
  \bibinfo {author} {\bibfnamefont {Chien-Chung}\ \bibnamefont {Hung}},
  \bibinfo {author} {\bibfnamefont {Kuei-Hung}\ \bibnamefont {Shen}}, \bibinfo
  {author} {\bibfnamefont {Ming-Jinn}\ \bibnamefont {Tsai}}, \ and\ \bibinfo
  {author} {\bibfnamefont {Ming-Jer}\ \bibnamefont {Kao}},\ }\bibfield  {title}
  {\enquote {\bibinfo {title} {{Reduction in critical current density for spin
  torque transfer switching with composite free layer}},}\ }\href {\doibase
  10.1063/1.2978097} {\bibfield  {journal} {\bibinfo  {journal} {Applied
  Physics Letters}\ }\textbf {\bibinfo {volume} {93}},\ \bibinfo {pages}
  {092504} (\bibinfo {year} {2008})}\BibitemShut {NoStop}%
\bibitem [{\citenamefont {Bosu}\ \emph {et~al.}(2016)\citenamefont {Bosu},
  \citenamefont {Sepehri-Amin}, \citenamefont {Sakuraba}, \citenamefont
  {Hayashi}, \citenamefont {Abert}, \citenamefont {Suess}, \citenamefont
  {Schrefl},\ and\ \citenamefont {Hono}}]{Bosu2016}%
  \BibitemOpen
  \bibfield  {author} {\bibinfo {author} {\bibfnamefont {S.}~\bibnamefont
  {Bosu}}, \bibinfo {author} {\bibfnamefont {H.}~\bibnamefont {Sepehri-Amin}},
  \bibinfo {author} {\bibfnamefont {Y.}~\bibnamefont {Sakuraba}}, \bibinfo
  {author} {\bibfnamefont {M.}~\bibnamefont {Hayashi}}, \bibinfo {author}
  {\bibfnamefont {C.}~\bibnamefont {Abert}}, \bibinfo {author} {\bibfnamefont
  {D.}~\bibnamefont {Suess}}, \bibinfo {author} {\bibfnamefont
  {T.}~\bibnamefont {Schrefl}}, \ and\ \bibinfo {author} {\bibfnamefont
  {K.}~\bibnamefont {Hono}},\ }\bibfield  {title} {\enquote {\bibinfo {title}
  {{Reduction of critical current density for out-of-plane mode oscillation in
  a mag-flip spin torque oscillator using highly spin-polarized Co$_{2}$
  Fe(Ga$_{0.5}$ Ge$_{0.5}$ ) spin injection layer}},}\ }\href {\doibase
  10.1063/1.4942373} {\bibfield  {journal} {\bibinfo  {journal} {Applied
  Physics Letters}\ }\textbf {\bibinfo {volume} {108}},\ \bibinfo {pages}
  {072403} (\bibinfo {year} {2016})}\BibitemShut {NoStop}%
\bibitem [{\citenamefont {Suess}\ \emph {et~al.}(2017)\citenamefont {Suess},
  \citenamefont {Vogler}, \citenamefont {Bruckner}, \citenamefont
  {Sepehri-Amin},\ and\ \citenamefont {Abert}}]{Suess2017}%
  \BibitemOpen
  \bibfield  {author} {\bibinfo {author} {\bibfnamefont {D.}~\bibnamefont
  {Suess}}, \bibinfo {author} {\bibfnamefont {C.}~\bibnamefont {Vogler}},
  \bibinfo {author} {\bibfnamefont {F.}~\bibnamefont {Bruckner}}, \bibinfo
  {author} {\bibfnamefont {H.}~\bibnamefont {Sepehri-Amin}}, \ and\ \bibinfo
  {author} {\bibfnamefont {C.}~\bibnamefont {Abert}},\ }\bibfield  {title}
  {\enquote {\bibinfo {title} {{Significant reduction of critical currents in
  MRAM designs using dual free layer with perpendicular and in-plane
  anisotropy}},}\ }\href {\doibase 10.1063/1.4987140} {\bibfield  {journal}
  {\bibinfo  {journal} {Applied Physics Letters}\ }\textbf {\bibinfo {volume}
  {110}},\ \bibinfo {pages} {252408} (\bibinfo {year} {2017})}\BibitemShut
  {NoStop}%
\bibitem [{\citenamefont {Weisheit}\ \emph {et~al.}(2007)\citenamefont
  {Weisheit}, \citenamefont {Fahler}, \citenamefont {Marty}, \citenamefont
  {Souche}, \citenamefont {Poinsignon},\ and\ \citenamefont
  {Givord}}]{Weisheit2007}%
  \BibitemOpen
  \bibfield  {author} {\bibinfo {author} {\bibfnamefont {M.}~\bibnamefont
  {Weisheit}}, \bibinfo {author} {\bibfnamefont {S.}~\bibnamefont {Fahler}},
  \bibinfo {author} {\bibfnamefont {A.}~\bibnamefont {Marty}}, \bibinfo
  {author} {\bibfnamefont {Y.}~\bibnamefont {Souche}}, \bibinfo {author}
  {\bibfnamefont {C.}~\bibnamefont {Poinsignon}}, \ and\ \bibinfo {author}
  {\bibfnamefont {D.}~\bibnamefont {Givord}},\ }\bibfield  {title} {\enquote
  {\bibinfo {title} {{Electric Field-Induced Modification of Magnetism in
  Thin-Film Ferromagnets}},}\ }\href@noop {} {\bibfield  {journal} {\bibinfo
  {journal} {Science}\ }\textbf {\bibinfo {volume} {315}},\ \bibinfo {pages}
  {349} (\bibinfo {year} {2007})}\BibitemShut {NoStop}%
\bibitem [{\citenamefont {Maruyama}\ \emph {et~al.}(2009)\citenamefont
  {Maruyama}, \citenamefont {Shiota}, \citenamefont {Nozaki}, \citenamefont
  {Ohta}, \citenamefont {Toda}, \citenamefont {Mizuguchi}, \citenamefont
  {Tulapurkar}, \citenamefont {Shinjo}, \citenamefont {Shiraishi},
  \citenamefont {Mizukami}, \citenamefont {Ando},\ and\ \citenamefont
  {Suzuki}}]{Maruyama2009}%
  \BibitemOpen
  \bibfield  {author} {\bibinfo {author} {\bibfnamefont {T.}~\bibnamefont
  {Maruyama}}, \bibinfo {author} {\bibfnamefont {Y.}~\bibnamefont {Shiota}},
  \bibinfo {author} {\bibfnamefont {T.}~\bibnamefont {Nozaki}}, \bibinfo
  {author} {\bibfnamefont {K.}~\bibnamefont {Ohta}}, \bibinfo {author}
  {\bibfnamefont {N.}~\bibnamefont {Toda}}, \bibinfo {author} {\bibfnamefont
  {M.}~\bibnamefont {Mizuguchi}}, \bibinfo {author} {\bibfnamefont {A.~A.}\
  \bibnamefont {Tulapurkar}}, \bibinfo {author} {\bibfnamefont
  {T.}~\bibnamefont {Shinjo}}, \bibinfo {author} {\bibfnamefont
  {M.}~\bibnamefont {Shiraishi}}, \bibinfo {author} {\bibfnamefont
  {S.}~\bibnamefont {Mizukami}}, \bibinfo {author} {\bibfnamefont
  {Y.}~\bibnamefont {Ando}}, \ and\ \bibinfo {author} {\bibfnamefont
  {Y.}~\bibnamefont {Suzuki}},\ }\bibfield  {title} {\enquote {\bibinfo {title}
  {{Large voltage-induced magnetic anisotropy change in a few atomic layers of
  iron}},}\ }\href@noop {} {\bibfield  {journal} {\bibinfo  {journal} {Nature
  Nanotechnology}\ }\textbf {\bibinfo {volume} {4}},\ \bibinfo {pages} {158}
  (\bibinfo {year} {2009})}\BibitemShut {NoStop}%
\bibitem [{\citenamefont {Nozaki}\ \emph {et~al.}(2010)\citenamefont {Nozaki},
  \citenamefont {Shiota}, \citenamefont {Shiraishi}, \citenamefont {Shinjo},\
  and\ \citenamefont {Suzuki}}]{Nozaki2010}%
  \BibitemOpen
  \bibfield  {author} {\bibinfo {author} {\bibfnamefont {T.}~\bibnamefont
  {Nozaki}}, \bibinfo {author} {\bibfnamefont {Y.}~\bibnamefont {Shiota}},
  \bibinfo {author} {\bibfnamefont {M.}~\bibnamefont {Shiraishi}}, \bibinfo
  {author} {\bibfnamefont {T.}~\bibnamefont {Shinjo}}, \ and\ \bibinfo {author}
  {\bibfnamefont {Y.}~\bibnamefont {Suzuki}},\ }\bibfield  {title} {\enquote
  {\bibinfo {title} {{Voltage-induced perpendicular magnetic anisotropy change
  in magnetic tunnel junctions}},}\ }\href@noop {} {\bibfield  {journal}
  {\bibinfo  {journal} {Applied Physics Letters}\ }\textbf {\bibinfo {volume}
  {96}},\ \bibinfo {pages} {3} (\bibinfo {year} {2010})}\BibitemShut {NoStop}%
\bibitem [{\citenamefont {Shiota}\ \emph {et~al.}(2011)\citenamefont {Shiota},
  \citenamefont {Nozaki}, \citenamefont {Bonell}, \citenamefont {Murakami},
  \citenamefont {Shinjo},\ and\ \citenamefont {Suzuki}}]{Shiota2011}%
  \BibitemOpen
  \bibfield  {author} {\bibinfo {author} {\bibfnamefont {Yoichi}\ \bibnamefont
  {Shiota}}, \bibinfo {author} {\bibfnamefont {Takayuki}\ \bibnamefont
  {Nozaki}}, \bibinfo {author} {\bibfnamefont {Fr{\'{e}}d{\'{e}}ric}\
  \bibnamefont {Bonell}}, \bibinfo {author} {\bibfnamefont {Shinichi}\
  \bibnamefont {Murakami}}, \bibinfo {author} {\bibfnamefont {Teruya}\
  \bibnamefont {Shinjo}}, \ and\ \bibinfo {author} {\bibfnamefont {Yoshishige}\
  \bibnamefont {Suzuki}},\ }\bibfield  {title} {\enquote {\bibinfo {title}
  {{Induction of coherent magnetization switching in a few atomic layers of
  FeCo using voltage pulses}},}\ }\href@noop {} {\bibfield  {journal} {\bibinfo
   {journal} {Nature Materials}\ }\textbf {\bibinfo {volume} {11}},\ \bibinfo
  {pages} {39} (\bibinfo {year} {2011})}\BibitemShut {NoStop}%
\bibitem [{\citenamefont {Nozaki}\ \emph {et~al.}(2014)\citenamefont {Nozaki},
  \citenamefont {Arai}, \citenamefont {Yakushiji}, \citenamefont {Tamaru},
  \citenamefont {Kubota}, \citenamefont {Imamura}, \citenamefont {Fukushima},\
  and\ \citenamefont {Yuasa}}]{Nozaki2014}%
  \BibitemOpen
  \bibfield  {author} {\bibinfo {author} {\bibfnamefont {Takayuki}\
  \bibnamefont {Nozaki}}, \bibinfo {author} {\bibfnamefont {Hiroko}\
  \bibnamefont {Arai}}, \bibinfo {author} {\bibfnamefont {Kay}\ \bibnamefont
  {Yakushiji}}, \bibinfo {author} {\bibfnamefont {Shingo}\ \bibnamefont
  {Tamaru}}, \bibinfo {author} {\bibfnamefont {Hitoshi}\ \bibnamefont
  {Kubota}}, \bibinfo {author} {\bibfnamefont {Hiroshi}\ \bibnamefont
  {Imamura}}, \bibinfo {author} {\bibfnamefont {Akio}\ \bibnamefont
  {Fukushima}}, \ and\ \bibinfo {author} {\bibfnamefont {Shinji}\ \bibnamefont
  {Yuasa}},\ }\bibfield  {title} {\enquote {\bibinfo {title} {{Magnetization
  switching assisted by high-frequency-voltage-induced ferromagnetic
  resonance}},}\ }\href@noop {} {\bibfield  {journal} {\bibinfo  {journal}
  {Applied Physics Express}\ }\textbf {\bibinfo {volume} {7}},\ \bibinfo
  {pages} {093005} (\bibinfo {year} {2014})}\BibitemShut {NoStop}%
\bibitem [{\citenamefont {Lin}\ \emph {et~al.}(2014)\citenamefont {Lin},
  \citenamefont {Chang}, \citenamefont {Tsai}, \citenamefont {Shieh},\ and\
  \citenamefont {Lo}}]{Lin2014}%
  \BibitemOpen
  \bibfield  {author} {\bibinfo {author} {\bibfnamefont {Wen~Chin}\
  \bibnamefont {Lin}}, \bibinfo {author} {\bibfnamefont {Po~Chun}\ \bibnamefont
  {Chang}}, \bibinfo {author} {\bibfnamefont {Cheng~Jui}\ \bibnamefont {Tsai}},
  \bibinfo {author} {\bibfnamefont {Tsung~Chun}\ \bibnamefont {Shieh}}, \ and\
  \bibinfo {author} {\bibfnamefont {Fang~Yuh}\ \bibnamefont {Lo}},\ }\bibfield
  {title} {\enquote {\bibinfo {title} {{Voltage-induced reversible changes in
  the magnetic coercivity of Fe/ZnO heterostructures}},}\ }\href@noop {}
  {\bibfield  {journal} {\bibinfo  {journal} {Applied Physics Letters}\
  }\textbf {\bibinfo {volume} {104}},\ \bibinfo {pages} {1} (\bibinfo {year}
  {2014})}\BibitemShut {NoStop}%
\bibitem [{\citenamefont {Amiri}\ \emph {et~al.}(2015)\citenamefont {Amiri},
  \citenamefont {Alzate}, \citenamefont {Cai}, \citenamefont {Ebrahimi},
  \citenamefont {Hu}, \citenamefont {Wong}, \citenamefont {Gr{\`{e}}zes},
  \citenamefont {Lee}, \citenamefont {Yu}, \citenamefont {Li}, \citenamefont
  {Akyol}, \citenamefont {Shao}, \citenamefont {Katine}, \citenamefont
  {Langer}, \citenamefont {Ocker},\ and\ \citenamefont {Wang}}]{Amiri2015}%
  \BibitemOpen
  \bibfield  {author} {\bibinfo {author} {\bibfnamefont {Pedram~Khalili}\
  \bibnamefont {Amiri}}, \bibinfo {author} {\bibfnamefont {Juan~G.}\
  \bibnamefont {Alzate}}, \bibinfo {author} {\bibfnamefont {Xue~Qing}\
  \bibnamefont {Cai}}, \bibinfo {author} {\bibfnamefont {Farbod}\ \bibnamefont
  {Ebrahimi}}, \bibinfo {author} {\bibfnamefont {Qi}~\bibnamefont {Hu}},
  \bibinfo {author} {\bibfnamefont {Kin}\ \bibnamefont {Wong}}, \bibinfo
  {author} {\bibfnamefont {C{\'{e}}cile}\ \bibnamefont {Gr{\`{e}}zes}},
  \bibinfo {author} {\bibfnamefont {Hochul}\ \bibnamefont {Lee}}, \bibinfo
  {author} {\bibfnamefont {Guoqiang}\ \bibnamefont {Yu}}, \bibinfo {author}
  {\bibfnamefont {Xiang}\ \bibnamefont {Li}}, \bibinfo {author} {\bibfnamefont
  {Mustafa}\ \bibnamefont {Akyol}}, \bibinfo {author} {\bibfnamefont {Qiming}\
  \bibnamefont {Shao}}, \bibinfo {author} {\bibfnamefont {Jordan~A.}\
  \bibnamefont {Katine}}, \bibinfo {author} {\bibfnamefont {J{\"{u}}rgen}\
  \bibnamefont {Langer}}, \bibinfo {author} {\bibfnamefont {Berthold}\
  \bibnamefont {Ocker}}, \ and\ \bibinfo {author} {\bibfnamefont {Kang~L.}\
  \bibnamefont {Wang}},\ }\bibfield  {title} {\enquote {\bibinfo {title}
  {{Electric-Field-Controlled Magnetoelectric RAM: Progress, Challenges, and
  Scaling}},}\ }\href@noop {} {\bibfield  {journal} {\bibinfo  {journal} {IEEE
  Transactions on Magnetics}\ }\textbf {\bibinfo {volume} {51}},\ \bibinfo
  {pages} {1} (\bibinfo {year} {2015})}\BibitemShut {NoStop}%
\bibitem [{\citenamefont {Kanai}\ \emph {et~al.}(2016)\citenamefont {Kanai},
  \citenamefont {Matsukura},\ and\ \citenamefont {Ohno}}]{Kanai2016}%
  \BibitemOpen
  \bibfield  {author} {\bibinfo {author} {\bibfnamefont {S.}~\bibnamefont
  {Kanai}}, \bibinfo {author} {\bibfnamefont {F.}~\bibnamefont {Matsukura}}, \
  and\ \bibinfo {author} {\bibfnamefont {H.}~\bibnamefont {Ohno}},\ }\bibfield
  {title} {\enquote {\bibinfo {title} {{Electric-field-induced magnetization
  switching in CoFeB/MgO magnetic tunnel junctions with high junction
  resistance}},}\ }\href {\doibase 10.1063/1.4948763} {\bibfield  {journal}
  {\bibinfo  {journal} {Applied Physics Letters}\ }\textbf {\bibinfo {volume}
  {108}},\ \bibinfo {pages} {2014} (\bibinfo {year} {2016})}\BibitemShut
  {NoStop}%
\bibitem [{\citenamefont {Grezes}\ \emph {et~al.}(2016)\citenamefont {Grezes},
  \citenamefont {Ebrahimi}, \citenamefont {Alzate}, \citenamefont {Cai},
  \citenamefont {Katine}, \citenamefont {Langer}, \citenamefont {Ocker},
  \citenamefont {{Khalili Amiri}},\ and\ \citenamefont {Wang}}]{Grezes2016}%
  \BibitemOpen
  \bibfield  {author} {\bibinfo {author} {\bibfnamefont {C.}~\bibnamefont
  {Grezes}}, \bibinfo {author} {\bibfnamefont {F.}~\bibnamefont {Ebrahimi}},
  \bibinfo {author} {\bibfnamefont {J.~G.}\ \bibnamefont {Alzate}}, \bibinfo
  {author} {\bibfnamefont {X.}~\bibnamefont {Cai}}, \bibinfo {author}
  {\bibfnamefont {J.~A.}\ \bibnamefont {Katine}}, \bibinfo {author}
  {\bibfnamefont {J.}~\bibnamefont {Langer}}, \bibinfo {author} {\bibfnamefont
  {B.}~\bibnamefont {Ocker}}, \bibinfo {author} {\bibfnamefont
  {P.}~\bibnamefont {{Khalili Amiri}}}, \ and\ \bibinfo {author} {\bibfnamefont
  {K.~L.}\ \bibnamefont {Wang}},\ }\bibfield  {title} {\enquote {\bibinfo
  {title} {{Ultra-low switching energy and scaling in electric-field-controlled
  nanoscale magnetic tunnel junctions with high resistance-area product}},}\
  }\href {\doibase 10.1063/1.4939446} {\bibfield  {journal} {\bibinfo
  {journal} {Applied Physics Letters}\ }\textbf {\bibinfo {volume} {108}},\
  \bibinfo {pages} {3} (\bibinfo {year} {2016})}\BibitemShut {NoStop}%
\bibitem [{\citenamefont {Munira}\ \emph {et~al.}(2016)\citenamefont {Munira},
  \citenamefont {Pandey}, \citenamefont {Kula},\ and\ \citenamefont
  {Sandhu}}]{Munira2016}%
  \BibitemOpen
  \bibfield  {author} {\bibinfo {author} {\bibfnamefont {Kamaram}\ \bibnamefont
  {Munira}}, \bibinfo {author} {\bibfnamefont {Sumeet~C.}\ \bibnamefont
  {Pandey}}, \bibinfo {author} {\bibfnamefont {Witold}\ \bibnamefont {Kula}}, \
  and\ \bibinfo {author} {\bibfnamefont {Gurtej~S.}\ \bibnamefont {Sandhu}},\
  }\bibfield  {title} {\enquote {\bibinfo {title} {{Voltage-controlled
  magnetization switching in MRAMs in conjunction with spin-transfer torque and
  applied magnetic field}},}\ }\href@noop {} {\bibfield  {journal} {\bibinfo
  {journal} {Journal of Applied Physics}\ }\textbf {\bibinfo {volume} {120}},\
  \bibinfo {pages} {203902} (\bibinfo {year} {2016})}\BibitemShut {NoStop}%
\bibitem [{\citenamefont {Nozaki}\ \emph {et~al.}(2016)\citenamefont {Nozaki},
  \citenamefont {Kozio{\l}-Rachwa{\l}}, \citenamefont {Skowro{\'{n}}ski},
  \citenamefont {Zayets}, \citenamefont {Shiota}, \citenamefont {Tamaru},
  \citenamefont {Kubota}, \citenamefont {Fukushima}, \citenamefont {Yuasa},\
  and\ \citenamefont {Suzuki}}]{Nozaki2016}%
  \BibitemOpen
  \bibfield  {author} {\bibinfo {author} {\bibfnamefont {Takayuki}\
  \bibnamefont {Nozaki}}, \bibinfo {author} {\bibfnamefont {Anna}\ \bibnamefont
  {Kozio{\l}-Rachwa{\l}}}, \bibinfo {author} {\bibfnamefont {Witold}\
  \bibnamefont {Skowro{\'{n}}ski}}, \bibinfo {author} {\bibfnamefont {Vadym}\
  \bibnamefont {Zayets}}, \bibinfo {author} {\bibfnamefont {Yoichi}\
  \bibnamefont {Shiota}}, \bibinfo {author} {\bibfnamefont {Shingo}\
  \bibnamefont {Tamaru}}, \bibinfo {author} {\bibfnamefont {Hitoshi}\
  \bibnamefont {Kubota}}, \bibinfo {author} {\bibfnamefont {Akio}\ \bibnamefont
  {Fukushima}}, \bibinfo {author} {\bibfnamefont {Shinji}\ \bibnamefont
  {Yuasa}}, \ and\ \bibinfo {author} {\bibfnamefont {Yoshishige}\ \bibnamefont
  {Suzuki}},\ }\bibfield  {title} {\enquote {\bibinfo {title} {{Large
  Voltage-Induced Changes in the Perpendicular Magnetic Anisotropy of an
  MgO-Based Tunnel Junction with an Ultrathin Fe Layer}},}\ }\href@noop {}
  {\bibfield  {journal} {\bibinfo  {journal} {Physical Review Applied}\
  }\textbf {\bibinfo {volume} {5}},\ \bibinfo {pages} {044006} (\bibinfo {year}
  {2016})}\BibitemShut {NoStop}%
\bibitem [{\citenamefont {Shiota}\ \emph {et~al.}(2016)\citenamefont {Shiota},
  \citenamefont {Nozaki}, \citenamefont {Tamaru}, \citenamefont {Yakushiji},
  \citenamefont {Kubota}, \citenamefont {Fukushima}, \citenamefont {Yuasa},\
  and\ \citenamefont {Suzuki}}]{Shiota2016}%
  \BibitemOpen
  \bibfield  {author} {\bibinfo {author} {\bibfnamefont {Yoichi}\ \bibnamefont
  {Shiota}}, \bibinfo {author} {\bibfnamefont {Takayuki}\ \bibnamefont
  {Nozaki}}, \bibinfo {author} {\bibfnamefont {Shingo}\ \bibnamefont {Tamaru}},
  \bibinfo {author} {\bibfnamefont {Kay}\ \bibnamefont {Yakushiji}}, \bibinfo
  {author} {\bibfnamefont {Hitoshi}\ \bibnamefont {Kubota}}, \bibinfo {author}
  {\bibfnamefont {Akio}\ \bibnamefont {Fukushima}}, \bibinfo {author}
  {\bibfnamefont {Shinji}\ \bibnamefont {Yuasa}}, \ and\ \bibinfo {author}
  {\bibfnamefont {Yoshishige}\ \bibnamefont {Suzuki}},\ }\bibfield  {title}
  {\enquote {\bibinfo {title} {{Evaluation of write error rate for
  voltage-driven dynamic magnetization switching in magnetic tunnel junctions
  with perpendicular magnetization}},}\ }\href@noop {} {\bibfield  {journal}
  {\bibinfo  {journal} {Applied Physics Express}\ }\textbf {\bibinfo {volume}
  {9}},\ \bibinfo {pages} {013001} (\bibinfo {year} {2016})}\BibitemShut
  {NoStop}%
\bibitem [{\citenamefont {Nozaki}\ \emph {et~al.}(2017)\citenamefont {Nozaki},
  \citenamefont {Kozio{\l}-Rachwa{\l}}, \citenamefont {Tsujikawa},
  \citenamefont {Shiota}, \citenamefont {Xu}, \citenamefont {Ohkubo},
  \citenamefont {Tsukahara}, \citenamefont {Miwa}, \citenamefont {Suzuki},
  \citenamefont {Tamaru}, \citenamefont {Kubota}, \citenamefont {Fukushima},
  \citenamefont {Hono}, \citenamefont {Shirai}, \citenamefont {Suzuki},\ and\
  \citenamefont {Yuasa}}]{Nozaki2017}%
  \BibitemOpen
  \bibfield  {author} {\bibinfo {author} {\bibfnamefont {Takayuki}\
  \bibnamefont {Nozaki}}, \bibinfo {author} {\bibfnamefont {Anna}\ \bibnamefont
  {Kozio{\l}-Rachwa{\l}}}, \bibinfo {author} {\bibfnamefont {Masahito}\
  \bibnamefont {Tsujikawa}}, \bibinfo {author} {\bibfnamefont {Yoichi}\
  \bibnamefont {Shiota}}, \bibinfo {author} {\bibfnamefont {Xiandong}\
  \bibnamefont {Xu}}, \bibinfo {author} {\bibfnamefont {Tadakatsu}\
  \bibnamefont {Ohkubo}}, \bibinfo {author} {\bibfnamefont {Takuya}\
  \bibnamefont {Tsukahara}}, \bibinfo {author} {\bibfnamefont {Shinji}\
  \bibnamefont {Miwa}}, \bibinfo {author} {\bibfnamefont {Motohiro}\
  \bibnamefont {Suzuki}}, \bibinfo {author} {\bibfnamefont {Shingo}\
  \bibnamefont {Tamaru}}, \bibinfo {author} {\bibfnamefont {Hitoshi}\
  \bibnamefont {Kubota}}, \bibinfo {author} {\bibfnamefont {Akio}\ \bibnamefont
  {Fukushima}}, \bibinfo {author} {\bibfnamefont {Kazuhiro}\ \bibnamefont
  {Hono}}, \bibinfo {author} {\bibfnamefont {Masafumi}\ \bibnamefont {Shirai}},
  \bibinfo {author} {\bibfnamefont {Yoshishige}\ \bibnamefont {Suzuki}}, \ and\
  \bibinfo {author} {\bibfnamefont {Shinji}\ \bibnamefont {Yuasa}},\ }\bibfield
   {title} {\enquote {\bibinfo {title} {{Highly efficient voltage control of
  spin and enhanced interfacial perpendicular magnetic anisotropy in
  iridium-doped Fe/MgO magnetic tunnel junctions}},}\ }\href@noop {} {\bibfield
   {journal} {\bibinfo  {journal} {NPG Asia Materials}\ }\textbf {\bibinfo
  {volume} {9}},\ \bibinfo {pages} {e451} (\bibinfo {year} {2017})}\BibitemShut
  {NoStop}%
\bibitem [{\citenamefont {Song}\ \emph {et~al.}(2017)\citenamefont {Song},
  \citenamefont {Cui}, \citenamefont {Li}, \citenamefont {Zhou},\ and\
  \citenamefont {Pan}}]{Song2017}%
  \BibitemOpen
  \bibfield  {author} {\bibinfo {author} {\bibfnamefont {Cheng}\ \bibnamefont
  {Song}}, \bibinfo {author} {\bibfnamefont {Bin}\ \bibnamefont {Cui}},
  \bibinfo {author} {\bibfnamefont {Fan}\ \bibnamefont {Li}}, \bibinfo {author}
  {\bibfnamefont {Xiangjun}\ \bibnamefont {Zhou}}, \ and\ \bibinfo {author}
  {\bibfnamefont {Feng}\ \bibnamefont {Pan}},\ }\bibfield  {title} {\enquote
  {\bibinfo {title} {{Recent progress in voltage control of magnetism:
  Materials, mechanisms, and performance}},}\ }\href {\doibase
  10.1016/j.pmatsci.2017.02.002} {\bibfield  {journal} {\bibinfo  {journal}
  {Progress in Materials Science}\ }\textbf {\bibinfo {volume} {87}},\ \bibinfo
  {pages} {33} (\bibinfo {year} {2017})}\BibitemShut {NoStop}%
\bibitem [{\citenamefont {Yamamoto}\ \emph {et~al.}(2018)\citenamefont
  {Yamamoto}, \citenamefont {Nozaki}, \citenamefont {Shiota}, \citenamefont
  {Imamura}, \citenamefont {Tamaru}, \citenamefont {Yakushiji}, \citenamefont
  {Kubota}, \citenamefont {Fukushima}, \citenamefont {Suzuki},\ and\
  \citenamefont {Yuasa}}]{Yamamoto2018}%
  \BibitemOpen
  \bibfield  {author} {\bibinfo {author} {\bibfnamefont {Tatsuya}\ \bibnamefont
  {Yamamoto}}, \bibinfo {author} {\bibfnamefont {Takayuki}\ \bibnamefont
  {Nozaki}}, \bibinfo {author} {\bibfnamefont {Yoichi}\ \bibnamefont {Shiota}},
  \bibinfo {author} {\bibfnamefont {Hiroshi}\ \bibnamefont {Imamura}}, \bibinfo
  {author} {\bibfnamefont {Shingo}\ \bibnamefont {Tamaru}}, \bibinfo {author}
  {\bibfnamefont {Kay}\ \bibnamefont {Yakushiji}}, \bibinfo {author}
  {\bibfnamefont {Hitoshi}\ \bibnamefont {Kubota}}, \bibinfo {author}
  {\bibfnamefont {Akio}\ \bibnamefont {Fukushima}}, \bibinfo {author}
  {\bibfnamefont {Yoshishige}\ \bibnamefont {Suzuki}}, \ and\ \bibinfo {author}
  {\bibfnamefont {Shinji}\ \bibnamefont {Yuasa}},\ }\bibfield  {title}
  {\enquote {\bibinfo {title} {{Thermally Induced Precession-Orbit Transition
  of Magnetization in Voltage-Driven Magnetization Switching}},}\ }\href
  {\doibase 10.1103/PhysRevApplied.10.024004} {\bibfield  {journal} {\bibinfo
  {journal} {Physical Review Applied}\ }\textbf {\bibinfo {volume} {10}},\
  \bibinfo {pages} {024004} (\bibinfo {year} {2018})}\BibitemShut {NoStop}%
\bibitem [{\citenamefont {Ikeura}\ \emph {et~al.}(2018)\citenamefont {Ikeura},
  \citenamefont {Nozaki}, \citenamefont {Shiota}, \citenamefont {Yamamoto},
  \citenamefont {Imamura}, \citenamefont {Kubota}, \citenamefont {Fukushima},
  \citenamefont {Suzuki},\ and\ \citenamefont {Yuasa}}]{Ikeura2018}%
  \BibitemOpen
  \bibfield  {author} {\bibinfo {author} {\bibfnamefont {T.}~\bibnamefont
  {Ikeura}}, \bibinfo {author} {\bibfnamefont {T.}~\bibnamefont {Nozaki}},
  \bibinfo {author} {\bibfnamefont {Y.}~\bibnamefont {Shiota}}, \bibinfo
  {author} {\bibfnamefont {T.}~\bibnamefont {Yamamoto}}, \bibinfo {author}
  {\bibfnamefont {H.}~\bibnamefont {Imamura}}, \bibinfo {author} {\bibfnamefont
  {H.}~\bibnamefont {Kubota}}, \bibinfo {author} {\bibfnamefont
  {A.}~\bibnamefont {Fukushima}}, \bibinfo {author} {\bibfnamefont
  {Y.}~\bibnamefont {Suzuki}}, \ and\ \bibinfo {author} {\bibfnamefont
  {S.}~\bibnamefont {Yuasa}},\ }\bibfield  {title} {\enquote {\bibinfo {title}
  {{Reduction in the write error rate of voltage-induced dynamic magnetization
  switching using the reverse bias method}},}\ }\href@noop {} {\bibfield
  {journal} {\bibinfo  {journal} {Japanese Journal of Applied Physics}\
  }\textbf {\bibinfo {volume} {57}} (\bibinfo {year} {2018})}\BibitemShut
  {NoStop}%
\bibitem [{\citenamefont {Matsumoto}\ \emph {et~al.}(2018)\citenamefont
  {Matsumoto}, \citenamefont {Nozaki}, \citenamefont {Yuasa},\ and\
  \citenamefont {Imamura}}]{Matsumoto2018}%
  \BibitemOpen
  \bibfield  {author} {\bibinfo {author} {\bibfnamefont {R.}~\bibnamefont
  {Matsumoto}}, \bibinfo {author} {\bibfnamefont {T.}~\bibnamefont {Nozaki}},
  \bibinfo {author} {\bibfnamefont {S.}~\bibnamefont {Yuasa}}, \ and\ \bibinfo
  {author} {\bibfnamefont {H.}~\bibnamefont {Imamura}},\ }\bibfield  {title}
  {\enquote {\bibinfo {title} {{Voltage-Induced Precessional Switching at
  Zero-Bias Magnetic Field in a Conically Magnetized Free Layer}},}\ }\href
  {\doibase 10.1103/PhysRevApplied.9.014026} {\bibfield  {journal} {\bibinfo
  {journal} {Physical Review Applied}\ }\textbf {\bibinfo {volume} {9}},\
  \bibinfo {pages} {014026} (\bibinfo {year} {2018})}\BibitemShut {NoStop}%
\bibitem [{\citenamefont {Miriyala}\ \emph {et~al.}(2019)\citenamefont
  {Miriyala}, \citenamefont {Fong},\ and\ \citenamefont {Liang}}]{Pavan2019}%
  \BibitemOpen
  \bibfield  {author} {\bibinfo {author} {\bibfnamefont {Venkata Pavan~Kumar}\
  \bibnamefont {Miriyala}}, \bibinfo {author} {\bibfnamefont {Xuanyao}\
  \bibnamefont {Fong}}, \ and\ \bibinfo {author} {\bibfnamefont {Gengchiau}\
  \bibnamefont {Liang}},\ }\bibfield  {title} {\enquote {\bibinfo {title}
  {{Influence of Size and Shape on the Performance of VCMA-Based MTJs}},}\
  }\href {\doibase 10.1109/TED.2018.2889112} {\bibfield  {journal} {\bibinfo
  {journal} {IEEE Transactions on Electron Devices}\ }\textbf {\bibinfo
  {volume} {66}},\ \bibinfo {pages} {944} (\bibinfo {year} {2019})}\BibitemShut
  {NoStop}%
\bibitem [{\citenamefont {Yamamoto}\ \emph {et~al.}(2019)\citenamefont
  {Yamamoto}, \citenamefont {Nozaki}, \citenamefont {Imamura}, \citenamefont
  {Shiota}, \citenamefont {Ikeura}, \citenamefont {Tamaru}, \citenamefont
  {Yakushiji}, \citenamefont {Kubota}, \citenamefont {Fukushima}, \citenamefont
  {Suzuki},\ and\ \citenamefont {Yuasa}}]{Yamamoto2019}%
  \BibitemOpen
  \bibfield  {author} {\bibinfo {author} {\bibfnamefont {Tatsuya}\ \bibnamefont
  {Yamamoto}}, \bibinfo {author} {\bibfnamefont {Takayuki}\ \bibnamefont
  {Nozaki}}, \bibinfo {author} {\bibfnamefont {Hiroshi}\ \bibnamefont
  {Imamura}}, \bibinfo {author} {\bibfnamefont {Yoichi}\ \bibnamefont
  {Shiota}}, \bibinfo {author} {\bibfnamefont {Takuro}\ \bibnamefont {Ikeura}},
  \bibinfo {author} {\bibfnamefont {Shingo}\ \bibnamefont {Tamaru}}, \bibinfo
  {author} {\bibfnamefont {Kay}\ \bibnamefont {Yakushiji}}, \bibinfo {author}
  {\bibfnamefont {Hitoshi}\ \bibnamefont {Kubota}}, \bibinfo {author}
  {\bibfnamefont {Akio}\ \bibnamefont {Fukushima}}, \bibinfo {author}
  {\bibfnamefont {Yoshishige}\ \bibnamefont {Suzuki}}, \ and\ \bibinfo {author}
  {\bibfnamefont {Shinji}\ \bibnamefont {Yuasa}},\ }\bibfield  {title}
  {\enquote {\bibinfo {title} {{Write-Error Reduction of Voltage-Torque-Driven
  Magnetization Switching by a Controlled Voltage Pulse}},}\ }\href {\doibase
  10.1103/PhysRevApplied.11.014013} {\bibfield  {journal} {\bibinfo  {journal}
  {Physical Review Applied}\ }\textbf {\bibinfo {volume} {11}},\ \bibinfo
  {pages} {014013} (\bibinfo {year} {2019})}\BibitemShut {NoStop}%
\bibitem [{\citenamefont {Matsumoto}\ \emph {et~al.}(2019)\citenamefont
  {Matsumoto}, \citenamefont {Sato},\ and\ \citenamefont
  {Imamura}}]{Matsumoto2019}%
  \BibitemOpen
  \bibfield  {author} {\bibinfo {author} {\bibfnamefont {Rie}\ \bibnamefont
  {Matsumoto}}, \bibinfo {author} {\bibfnamefont {Tomoyuki}\ \bibnamefont
  {Sato}}, \ and\ \bibinfo {author} {\bibfnamefont {Hiroshi}\ \bibnamefont
  {Imamura}},\ }\bibfield  {title} {\enquote {\bibinfo {title}
  {{Voltage-induced switching with long tolerance of voltage-pulse duration in
  a perpendicularly magnetized free layer}},}\ }\href {\doibase
  10.7567/1882-0786/ab1349} {\bibfield  {journal} {\bibinfo  {journal} {Applied
  Physics Express}\ }\textbf {\bibinfo {volume} {12}},\ \bibinfo {pages}
  {053003} (\bibinfo {year} {2019})}\BibitemShut {NoStop}%
\bibitem [{\citenamefont {Duan}\ \emph {et~al.}(2008)\citenamefont {Duan},
  \citenamefont {Velev}, \citenamefont {Sabirianov}, \citenamefont {Zhu},
  \citenamefont {Chu}, \citenamefont {Jaswal},\ and\ \citenamefont
  {Tsymbal}}]{Duan2008}%
  \BibitemOpen
  \bibfield  {author} {\bibinfo {author} {\bibfnamefont {Chun-Gang}\
  \bibnamefont {Duan}}, \bibinfo {author} {\bibfnamefont {Julian~P.}\
  \bibnamefont {Velev}}, \bibinfo {author} {\bibfnamefont {R.~F.}\ \bibnamefont
  {Sabirianov}}, \bibinfo {author} {\bibfnamefont {Ziqiang}\ \bibnamefont
  {Zhu}}, \bibinfo {author} {\bibfnamefont {Junhao}\ \bibnamefont {Chu}},
  \bibinfo {author} {\bibfnamefont {S.~S.}\ \bibnamefont {Jaswal}}, \ and\
  \bibinfo {author} {\bibfnamefont {E.~Y.}\ \bibnamefont {Tsymbal}},\
  }\bibfield  {title} {\enquote {\bibinfo {title} {Surface magnetoelectric
  effect in ferromagnetic metal films},}\ }\href {\doibase
  10.1103/PhysRevLett.101.137201} {\bibfield  {journal} {\bibinfo  {journal}
  {Phys. Rev. Lett.}\ }\textbf {\bibinfo {volume} {101}},\ \bibinfo {pages}
  {137201} (\bibinfo {year} {2008})}\BibitemShut {NoStop}%
\bibitem [{\citenamefont {Nakamura}\ \emph {et~al.}(2009)\citenamefont
  {Nakamura}, \citenamefont {Shimabukuro}, \citenamefont {Fujiwara},
  \citenamefont {Akiyama}, \citenamefont {Ito},\ and\ \citenamefont
  {Freeman}}]{Nakamura2009}%
  \BibitemOpen
  \bibfield  {author} {\bibinfo {author} {\bibfnamefont {Kohji}\ \bibnamefont
  {Nakamura}}, \bibinfo {author} {\bibfnamefont {Riki}\ \bibnamefont
  {Shimabukuro}}, \bibinfo {author} {\bibfnamefont {Yuji}\ \bibnamefont
  {Fujiwara}}, \bibinfo {author} {\bibfnamefont {Toru}\ \bibnamefont
  {Akiyama}}, \bibinfo {author} {\bibfnamefont {Tomonori}\ \bibnamefont {Ito}},
  \ and\ \bibinfo {author} {\bibfnamefont {A~J}\ \bibnamefont {Freeman}},\
  }\bibfield  {title} {\enquote {\bibinfo {title} {{Giant Modification of the
  Magnetocrystalline Anisotropy in Transition-Metal Monolayers by an External
  Electric Field}},}\ }\href {\doibase 10.1103/PhysRevLett.102.187201}
  {\bibfield  {journal} {\bibinfo  {journal} {Physical Review Letters}\
  }\textbf {\bibinfo {volume} {102}},\ \bibinfo {pages} {187201} (\bibinfo
  {year} {2009})}\BibitemShut {NoStop}%
\bibitem [{\citenamefont {Tsujikawa}\ and\ \citenamefont
  {Oda}(2009)}]{Tsujikawa2009}%
  \BibitemOpen
  \bibfield  {author} {\bibinfo {author} {\bibfnamefont {Masahito}\
  \bibnamefont {Tsujikawa}}\ and\ \bibinfo {author} {\bibfnamefont {Tatsuki}\
  \bibnamefont {Oda}},\ }\bibfield  {title} {\enquote {\bibinfo {title}
  {{Finite electric field effects in the large perpendicular magnetic
  anisotropy surface Pt/Fe/Pt(001): A first-principles study}},}\ }\href
  {\doibase 10.1103/PhysRevLett.102.247203} {\bibfield  {journal} {\bibinfo
  {journal} {Physical Review Letters}\ }\textbf {\bibinfo {volume} {102}},\
  \bibinfo {pages} {247203} (\bibinfo {year} {2009})}\BibitemShut {NoStop}%
\bibitem [{\citenamefont {Niranjan}\ \emph {et~al.}(2010)\citenamefont
  {Niranjan}, \citenamefont {Duan}, \citenamefont {Jaswal},\ and\ \citenamefont
  {Tsymbal}}]{Niranjan2010}%
  \BibitemOpen
  \bibfield  {author} {\bibinfo {author} {\bibfnamefont {Manish~K.}\
  \bibnamefont {Niranjan}}, \bibinfo {author} {\bibfnamefont {Chun-Gang}\
  \bibnamefont {Duan}}, \bibinfo {author} {\bibfnamefont {Sitaram~S.}\
  \bibnamefont {Jaswal}}, \ and\ \bibinfo {author} {\bibfnamefont {Evgeny~Y.}\
  \bibnamefont {Tsymbal}},\ }\bibfield  {title} {\enquote {\bibinfo {title}
  {{Electric field effect on magnetization at the Fe/MgO(001) interface}},}\
  }\href {\doibase 10.1063/1.3443658} {\bibfield  {journal} {\bibinfo
  {journal} {Applied Physics Letters}\ }\textbf {\bibinfo {volume} {96}},\
  \bibinfo {pages} {222504} (\bibinfo {year} {2010})}\BibitemShut {NoStop}%
\bibitem [{\citenamefont {Miwa}\ \emph {et~al.}(2017)\citenamefont {Miwa},
  \citenamefont {Suzuki}, \citenamefont {Tsujikawa}, \citenamefont {Matsuda},
  \citenamefont {Nozaki}, \citenamefont {Tanaka}, \citenamefont {Tsukahara},
  \citenamefont {Nawaoka}, \citenamefont {Goto}, \citenamefont {Kotani},
  \citenamefont {Ohkubo}, \citenamefont {Bonell}, \citenamefont {Tamura},
  \citenamefont {Hono}, \citenamefont {Nakamura}, \citenamefont {Shirai},
  \citenamefont {Yuasa},\ and\ \citenamefont {Suzuki}}]{Miwa2017}%
  \BibitemOpen
  \bibfield  {author} {\bibinfo {author} {\bibfnamefont {Shinji}\ \bibnamefont
  {Miwa}}, \bibinfo {author} {\bibfnamefont {Motohiro}\ \bibnamefont {Suzuki}},
  \bibinfo {author} {\bibfnamefont {Masahito}\ \bibnamefont {Tsujikawa}},
  \bibinfo {author} {\bibfnamefont {Kensho}\ \bibnamefont {Matsuda}}, \bibinfo
  {author} {\bibfnamefont {Takayuki}\ \bibnamefont {Nozaki}}, \bibinfo {author}
  {\bibfnamefont {Kazuhito}\ \bibnamefont {Tanaka}}, \bibinfo {author}
  {\bibfnamefont {Takuya}\ \bibnamefont {Tsukahara}}, \bibinfo {author}
  {\bibfnamefont {Kohei}\ \bibnamefont {Nawaoka}}, \bibinfo {author}
  {\bibfnamefont {Minori}\ \bibnamefont {Goto}}, \bibinfo {author}
  {\bibfnamefont {Yoshinori}\ \bibnamefont {Kotani}}, \bibinfo {author}
  {\bibfnamefont {Tadakatsu}\ \bibnamefont {Ohkubo}}, \bibinfo {author}
  {\bibfnamefont {Fr{\'{e}}d{\'{e}}ric}\ \bibnamefont {Bonell}}, \bibinfo
  {author} {\bibfnamefont {Eiiti}\ \bibnamefont {Tamura}}, \bibinfo {author}
  {\bibfnamefont {Kazuhiro}\ \bibnamefont {Hono}}, \bibinfo {author}
  {\bibfnamefont {Tetsuya}\ \bibnamefont {Nakamura}}, \bibinfo {author}
  {\bibfnamefont {Masafumi}\ \bibnamefont {Shirai}}, \bibinfo {author}
  {\bibfnamefont {Shinji}\ \bibnamefont {Yuasa}}, \ and\ \bibinfo {author}
  {\bibfnamefont {Yoshishige}\ \bibnamefont {Suzuki}},\ }\bibfield  {title}
  {\enquote {\bibinfo {title} {{Voltage controlled interfacial magnetism
  through platinum orbits}},}\ }\href {\doibase 10.1038/ncomms15848} {\bibfield
   {journal} {\bibinfo  {journal} {Nature Communications}\ }\textbf {\bibinfo
  {volume} {8}},\ \bibinfo {pages} {15848} (\bibinfo {year}
  {2017})}\BibitemShut {NoStop}%
\bibitem [{\citenamefont {Worledge}\ \emph {et~al.}(2010)\citenamefont
  {Worledge}, \citenamefont {Hu}, \citenamefont {Trouilloud}, \citenamefont
  {Abraham}, \citenamefont {Brown}, \citenamefont {Gaidis}, \citenamefont
  {Nowak}, \citenamefont {O'Sullivan}, \citenamefont {Robertazzi},
  \citenamefont {Sun},\ and\ \citenamefont {Gallagher}}]{Worledge2010}%
  \BibitemOpen
  \bibfield  {author} {\bibinfo {author} {\bibfnamefont {D.~C.}\ \bibnamefont
  {Worledge}}, \bibinfo {author} {\bibfnamefont {G.}~\bibnamefont {Hu}},
  \bibinfo {author} {\bibfnamefont {P.~L.}\ \bibnamefont {Trouilloud}},
  \bibinfo {author} {\bibfnamefont {D.~W.}\ \bibnamefont {Abraham}}, \bibinfo
  {author} {\bibfnamefont {S.}~\bibnamefont {Brown}}, \bibinfo {author}
  {\bibfnamefont {M.~C.}\ \bibnamefont {Gaidis}}, \bibinfo {author}
  {\bibfnamefont {J.}~\bibnamefont {Nowak}}, \bibinfo {author} {\bibfnamefont
  {E.~J.}\ \bibnamefont {O'Sullivan}}, \bibinfo {author} {\bibfnamefont
  {R.~P.}\ \bibnamefont {Robertazzi}}, \bibinfo {author} {\bibfnamefont
  {J.~Z.}\ \bibnamefont {Sun}}, \ and\ \bibinfo {author} {\bibfnamefont
  {W.~J.}\ \bibnamefont {Gallagher}},\ }\bibfield  {title} {\enquote {\bibinfo
  {title} {{Switching distributions and write reliability of perpendicular spin
  torque MRAM}},}\ }in\ \href {\doibase 10.1109/IEDM.2010.5703349} {\emph
  {\bibinfo {booktitle} {2010 International Electron Devices Meeting}}}\
  (\bibinfo  {publisher} {IEEE},\ \bibinfo {year} {2010})\ p.\ \bibinfo {pages}
  {12.5.1}\BibitemShut {NoStop}%
\bibitem [{\citenamefont {Min}\ \emph {et~al.}(2010)\citenamefont {Min},
  \citenamefont {Chen}, \citenamefont {Beach}, \citenamefont {Jan},
  \citenamefont {Horng}, \citenamefont {Kula}, \citenamefont {Torng},
  \citenamefont {Tong}, \citenamefont {Zhong}, \citenamefont {Tang},
  \citenamefont {Wang}, \citenamefont {Chen}, \citenamefont {Sun},
  \citenamefont {Debrosse}, \citenamefont {Worledge}, \citenamefont {Maffitt},\
  and\ \citenamefont {Gallagher}}]{Min2010}%
  \BibitemOpen
  \bibfield  {author} {\bibinfo {author} {\bibfnamefont {Tai}\ \bibnamefont
  {Min}}, \bibinfo {author} {\bibfnamefont {Qiang}\ \bibnamefont {Chen}},
  \bibinfo {author} {\bibfnamefont {Robert}\ \bibnamefont {Beach}}, \bibinfo
  {author} {\bibfnamefont {Guenole}\ \bibnamefont {Jan}}, \bibinfo {author}
  {\bibfnamefont {Cheng}\ \bibnamefont {Horng}}, \bibinfo {author}
  {\bibfnamefont {Witold}\ \bibnamefont {Kula}}, \bibinfo {author}
  {\bibfnamefont {Terry}\ \bibnamefont {Torng}}, \bibinfo {author}
  {\bibfnamefont {Ruth}\ \bibnamefont {Tong}}, \bibinfo {author} {\bibfnamefont
  {Tom}\ \bibnamefont {Zhong}}, \bibinfo {author} {\bibfnamefont {Denny}\
  \bibnamefont {Tang}}, \bibinfo {author} {\bibfnamefont {Pokang}\ \bibnamefont
  {Wang}}, \bibinfo {author} {\bibfnamefont {Mao~Min}\ \bibnamefont {Chen}},
  \bibinfo {author} {\bibfnamefont {J.~Z.}\ \bibnamefont {Sun}}, \bibinfo
  {author} {\bibfnamefont {J.~K.}\ \bibnamefont {Debrosse}}, \bibinfo {author}
  {\bibfnamefont {D.~C.}\ \bibnamefont {Worledge}}, \bibinfo {author}
  {\bibfnamefont {T.~M.}\ \bibnamefont {Maffitt}}, \ and\ \bibinfo {author}
  {\bibfnamefont {W.~J.}\ \bibnamefont {Gallagher}},\ }\bibfield  {title}
  {\enquote {\bibinfo {title} {{A study of write margin of spin torque transfer
  magnetic random access memory technology}},}\ }\href {\doibase
  10.1109/TMAG.2010.2043069} {\bibfield  {journal} {\bibinfo  {journal} {IEEE
  Transactions on Magnetics}\ }\textbf {\bibinfo {volume} {46}},\ \bibinfo
  {pages} {2322} (\bibinfo {year} {2010})}\BibitemShut {NoStop}%
\bibitem [{\citenamefont {Nowak}\ \emph {et~al.}(2011)\citenamefont {Nowak},
  \citenamefont {Robertazzi}, \citenamefont {Sun}, \citenamefont {Hu},
  \citenamefont {Abraham}, \citenamefont {Trouilloud}, \citenamefont {Brown},
  \citenamefont {Gaidis}, \citenamefont {O'Sullivan}, \citenamefont
  {Gallagher},\ and\ \citenamefont {Worledge}}]{Nowak2011}%
  \BibitemOpen
  \bibfield  {author} {\bibinfo {author} {\bibfnamefont {J.~J.}\ \bibnamefont
  {Nowak}}, \bibinfo {author} {\bibfnamefont {R.~P.}\ \bibnamefont
  {Robertazzi}}, \bibinfo {author} {\bibfnamefont {J.~Z.}\ \bibnamefont {Sun}},
  \bibinfo {author} {\bibfnamefont {G.}~\bibnamefont {Hu}}, \bibinfo {author}
  {\bibfnamefont {David~W.}\ \bibnamefont {Abraham}}, \bibinfo {author}
  {\bibfnamefont {P.~L.}\ \bibnamefont {Trouilloud}}, \bibinfo {author}
  {\bibfnamefont {S.}~\bibnamefont {Brown}}, \bibinfo {author} {\bibfnamefont
  {M.~C.}\ \bibnamefont {Gaidis}}, \bibinfo {author} {\bibfnamefont {E.~J.}\
  \bibnamefont {O'Sullivan}}, \bibinfo {author} {\bibfnamefont {W.~J.}\
  \bibnamefont {Gallagher}}, \ and\ \bibinfo {author} {\bibfnamefont {D.~C.}\
  \bibnamefont {Worledge}},\ }\bibfield  {title} {\enquote {\bibinfo {title}
  {{Demonstration of ultralow bit error rates for spin-torque magnetic
  random-access memory with perpendicular magnetic anisotropy}},}\ }\href
  {\doibase 10.1109/LMAG.2011.2155625} {\bibfield  {journal} {\bibinfo
  {journal} {IEEE Magnetics Letters}\ }\textbf {\bibinfo {volume} {2}},\
  \bibinfo {pages} {2} (\bibinfo {year} {2011})}\BibitemShut {NoStop}%
\bibitem [{\citenamefont {Sun}\ \emph {et~al.}(2012)\citenamefont {Sun},
  \citenamefont {Liu}, \citenamefont {Min}, \citenamefont {Zheng},\ and\
  \citenamefont {Zhang}}]{Sun2012}%
  \BibitemOpen
  \bibfield  {author} {\bibinfo {author} {\bibfnamefont {Hongbin}\ \bibnamefont
  {Sun}}, \bibinfo {author} {\bibfnamefont {Chuanyin}\ \bibnamefont {Liu}},
  \bibinfo {author} {\bibfnamefont {Tai}\ \bibnamefont {Min}}, \bibinfo
  {author} {\bibfnamefont {Nanning}\ \bibnamefont {Zheng}}, \ and\ \bibinfo
  {author} {\bibfnamefont {Tong}\ \bibnamefont {Zhang}},\ }\bibfield  {title}
  {\enquote {\bibinfo {title} {{Architectural Exploration to Enable Sufficient
  MTJ Device Write Margin for STT-RAM Based Cache}},}\ }\href {\doibase
  10.1109/TMAG.2012.2193589} {\bibfield  {journal} {\bibinfo  {journal} {IEEE
  Transactions on Magnetics}\ }\textbf {\bibinfo {volume} {48}},\ \bibinfo
  {pages} {2346} (\bibinfo {year} {2012})}\BibitemShut {NoStop}%
\bibitem [{\citenamefont {Shiota}\ \emph {et~al.}(2017)\citenamefont {Shiota},
  \citenamefont {Nozaki}, \citenamefont {Tamaru}, \citenamefont {Yakushiji},
  \citenamefont {Kubota}, \citenamefont {Fukushima}, \citenamefont {Yuasa},\
  and\ \citenamefont {Suzuki}}]{Shiota2017}%
  \BibitemOpen
  \bibfield  {author} {\bibinfo {author} {\bibfnamefont {Yoichi}\ \bibnamefont
  {Shiota}}, \bibinfo {author} {\bibfnamefont {Takayuki}\ \bibnamefont
  {Nozaki}}, \bibinfo {author} {\bibfnamefont {Shingo}\ \bibnamefont {Tamaru}},
  \bibinfo {author} {\bibfnamefont {Kay}\ \bibnamefont {Yakushiji}}, \bibinfo
  {author} {\bibfnamefont {Hitoshi}\ \bibnamefont {Kubota}}, \bibinfo {author}
  {\bibfnamefont {Akio}\ \bibnamefont {Fukushima}}, \bibinfo {author}
  {\bibfnamefont {Shinji}\ \bibnamefont {Yuasa}}, \ and\ \bibinfo {author}
  {\bibfnamefont {Yoshishige}\ \bibnamefont {Suzuki}},\ }\bibfield  {title}
  {\enquote {\bibinfo {title} {{Reduction in write error rate of voltage-driven
  dynamic magnetization switching by improving thermal stability factor}},}\
  }\href {\doibase 10.1063/1.4990680} {\bibfield  {journal} {\bibinfo
  {journal} {Applied Physics Letters}\ }\textbf {\bibinfo {volume} {111}},\
  \bibinfo {pages} {2} (\bibinfo {year} {2017})}\BibitemShut {NoStop}%
\bibitem [{\citenamefont {Brown}(1963)}]{Brown1963}%
  \BibitemOpen
  \bibfield  {author} {\bibinfo {author} {\bibfnamefont {William~Fuller}\
  \bibnamefont {Brown}},\ }\bibfield  {title} {\enquote {\bibinfo {title}
  {{Thermal fluctuations of a single-domain particle}},}\ }\href {\doibase
  10.1103/PhysRev.130.1677} {\bibfield  {journal} {\bibinfo  {journal}
  {Physical Review}\ }\textbf {\bibinfo {volume} {130}},\ \bibinfo {pages}
  {1677} (\bibinfo {year} {1963})}\BibitemShut {NoStop}%
\bibitem [{\citenamefont {Callen}\ and\ \citenamefont
  {Welton}(1951)}]{Callen1951}%
  \BibitemOpen
  \bibfield  {author} {\bibinfo {author} {\bibfnamefont {Herbert~B}\
  \bibnamefont {Callen}}\ and\ \bibinfo {author} {\bibfnamefont {Theodore~A.}\
  \bibnamefont {Welton}},\ }\bibfield  {title} {\enquote {\bibinfo {title}
  {{Irreversibility and Generalized Noise}},}\ }\href {\doibase
  10.1103/PhysRev.83.34} {\bibfield  {journal} {\bibinfo  {journal} {Physical
  Review}\ }\textbf {\bibinfo {volume} {83}},\ \bibinfo {pages} {34} (\bibinfo
  {year} {1951})}\BibitemShut {NoStop}%
\bibitem [{\citenamefont {Callen}\ and\ \citenamefont
  {Greene}(1952)}]{Callen1952a}%
  \BibitemOpen
  \bibfield  {author} {\bibinfo {author} {\bibfnamefont {Herbert~B.}\
  \bibnamefont {Callen}}\ and\ \bibinfo {author} {\bibfnamefont {Richard~F.}\
  \bibnamefont {Greene}},\ }\bibfield  {title} {\enquote {\bibinfo {title} {{On
  a Theorem of Irreversible Thermodynamics}},}\ }\href {\doibase
  10.1103/PhysRev.86.702} {\bibfield  {journal} {\bibinfo  {journal} {Physical
  Review}\ }\textbf {\bibinfo {volume} {86}},\ \bibinfo {pages} {702} (\bibinfo
  {year} {1952})}\BibitemShut {NoStop}%
\bibitem [{\citenamefont {Callen}\ \emph {et~al.}(1952)\citenamefont {Callen},
  \citenamefont {Barasch},\ and\ \citenamefont {Jackson}}]{Callen1952b}%
  \BibitemOpen
  \bibfield  {author} {\bibinfo {author} {\bibfnamefont {Herbert~B.}\
  \bibnamefont {Callen}}, \bibinfo {author} {\bibfnamefont {Murray~L.}\
  \bibnamefont {Barasch}}, \ and\ \bibinfo {author} {\bibfnamefont {Julius~L.}\
  \bibnamefont {Jackson}},\ }\bibfield  {title} {\enquote {\bibinfo {title}
  {{Statistical Mechanics of Irreversibility}},}\ }\href {\doibase
  10.1103/PhysRev.88.1382} {\bibfield  {journal} {\bibinfo  {journal} {Physical
  Review}\ }\textbf {\bibinfo {volume} {88}},\ \bibinfo {pages} {1382}
  (\bibinfo {year} {1952})}\BibitemShut {NoStop}%
\bibitem [{\citenamefont {Greene}\ and\ \citenamefont
  {Callen}(1952)}]{Greene1952}%
  \BibitemOpen
  \bibfield  {author} {\bibinfo {author} {\bibfnamefont {Richard~F.}\
  \bibnamefont {Greene}}\ and\ \bibinfo {author} {\bibfnamefont {Herbert~B.}\
  \bibnamefont {Callen}},\ }\bibfield  {title} {\enquote {\bibinfo {title} {{On
  a Theorem of Irreversible Thermodynamics. II}},}\ }\href {\doibase
  10.1103/PhysRev.88.1387} {\bibfield  {journal} {\bibinfo  {journal} {Physical
  Review}\ }\textbf {\bibinfo {volume} {88}},\ \bibinfo {pages} {1387}
  (\bibinfo {year} {1952})}\BibitemShut {NoStop}%
\bibitem [{\citenamefont {Stiles}\ and\ \citenamefont
  {Miltat}(2005)}]{Stiles2005}%
  \BibitemOpen
  \bibfield  {author} {\bibinfo {author} {\bibfnamefont {Mark~D}\ \bibnamefont
  {Stiles}}\ and\ \bibinfo {author} {\bibfnamefont {Jacques}\ \bibnamefont
  {Miltat}},\ }\bibfield  {title} {\enquote {\bibinfo {title} {{Spin-Transfer
  Torque and Dynamics}},}\ }in\ \href {\doibase 10.1007/10938171_7} {\emph
  {\bibinfo {booktitle} {Spin Dynamics in Confined Magnetic Structures III}}},\
  Vol.\ \bibinfo {volume} {101}\ (\bibinfo  {publisher} {Springer Berlin
  Heidelberg},\ \bibinfo {year} {2005})\ p.\ \bibinfo {pages} {225}\BibitemShut
  {NoStop}%
\bibitem [{\citenamefont {Apalkov}\ and\ \citenamefont
  {Visscher}(2005)}]{Apalkov2005b}%
  \BibitemOpen
  \bibfield  {author} {\bibinfo {author} {\bibfnamefont {D.~M.}\ \bibnamefont
  {Apalkov}}\ and\ \bibinfo {author} {\bibfnamefont {P.~B.}\ \bibnamefont
  {Visscher}},\ }\bibfield  {title} {\enquote {\bibinfo {title} {{Spin-torque
  switching: Fokker-Planck rate calculation}},}\ }\href {\doibase
  10.1103/PhysRevB.72.180405} {\bibfield  {journal} {\bibinfo  {journal}
  {Physical Review B}\ }\textbf {\bibinfo {volume} {72}},\ \bibinfo {pages} {180405(R)}
  (\bibinfo {year} {2005})}\BibitemShut {NoStop}%
\bibitem [{\citenamefont {Tzoufras}(2018)}]{Tzoufras2018}%
  \BibitemOpen
  \bibfield  {author} {\bibinfo {author} {\bibfnamefont {M}~\bibnamefont
  {Tzoufras}},\ }\bibfield  {title} {\enquote {\bibinfo {title} {{Switching
  probability of all-perpendicular spin valve nanopillars}},}\ }\href {\doibase
  10.1063/1.5003832} {\bibfield  {journal} {\bibinfo  {journal} {AIP Advances}\
  }\textbf {\bibinfo {volume} {8}},\ \bibinfo {pages} {56002} (\bibinfo {year}
  {2018})}\BibitemShut {NoStop}%
\end{thebibliography}
\end{document}